\documentclass[pra,twocolumn,showpacs,aps,superscriptaddress]{revtex4-1}
 
\usepackage{amsmath}
\usepackage{amssymb}
\usepackage{amsmath}
\usepackage{txfonts}
\usepackage{graphicx}
\usepackage{epsfig}
\usepackage[T1]{fontenc}
\usepackage[latin9]{inputenc}
\usepackage{times}
\usepackage{color} 
\usepackage{xspace}
\usepackage{amssymb,amsmath}
\usepackage{amsbsy}
\usepackage{braket}
\usepackage{tabularx}
\usepackage{mathtools}          
\usepackage{capt-of} 

\DeclareMathOperator{\sech}{sech}

\usepackage{ifpdf}
\ifpdf
\usepackage{epstopdf}
\fi

\usepackage[unicode]{hyperref}
\hypersetup{
	pdftitle={spinnorBEC},
	pdfauthor={Yu},
	pdfsubject={Vortexlines},
	colorlinks=true,
	linkcolor=blue,
	citecolor=blue,
	filecolor=black,
	urlcolor=blue
}
\def\urlprefix{}
\def\url#1{}

\usepackage{bm}
\usepackage{color}

\newcommand{\be}{\begin{equation}}

\newcommand{\ee}{\end{equation}}

\newcommand{\bea}{\begin{eqnarray}}

\newcommand{\eea}{\end{eqnarray}}

\begin{document}

\title{Absence of the breakdown of ferrodark solitons exhibiting a snake instability}

\author{Xiaoquan Yu}
\email{xqyu@gscaep.ac.cn}
\affiliation{Graduate School of China Academy of Engineering Physics, Beijing 100193, China}
\affiliation{Department of Physics, Centre for Quantum Science, and Dodd-Walls Centre for Photonic and Quantum Technologies, University of Otago, Dunedin, New Zealand}
\author{P.~B.~Blakie}
\affiliation{Department of Physics, Centre for Quantum Science, and Dodd-Walls Centre for Photonic and Quantum Technologies, University of Otago, Dunedin, New Zealand}

\begin{abstract} 
We investigate the dynamical stability and real time dynamics of the two-types of ferrodark solitons (FDSs) which occur as  topological magnetic domain walls in the easy-plane phase of a quasi-two-dimensional (2D) ferromagnetic spin-1 Bose-Einstein condensate. The type-I  FDS  has positive inertial mass and exhibits a single dynamical instability  that generates in plane spin winding, causing polar-core spin vortex dipoles. The positive inertial mass leads to the elastic oscillations of the soliton under transverse perturbations.  
The type-II FDS has negative inertial mass and exhibits a snake instability and a spin-twist instability, with the latter involving the generation of  out of plane spin winding.  Distinct from the normal dynamics of negative mass solitons under long wave length transverse perturbations, the snake instability does not lead to the type-II FDS breaking down.  Instead, segments of the type-II FDS convert to type-I and mass vortex dipoles are produced. The resulting hybridized-chain of the two soliton types and vortices  appear as a stable vortex-domain wall composite excitation  and  exhibits complex 2D dynamics at long times, while  the  vortices remain confined and  the topological structure of a magnetic domain wall persists.   

\end{abstract}

\maketitle

\textit{Introduction---}
Dark solitons in a Bose-Einstein condensate (BEC) are unstable 
against transverse perturbations in high dimensions ($d>1$)~\cite{kuznetsov1988instability,Shlyapnikov1999}, known as the snake instability. Such an instability is a general phenomenon and exists for topological solitons in various systems~\cite{Pitaevskii2011,Gallemi2019,MDQu2017, Sophie,MDQu2016,MSexp1,MSexp2}. 
It has been demonstrated that the presence of the snake instability occurs when the inertial mass, which appears in the Newton equation describing one-dimensional (1D) soliton motion in superfluids~\cite{Pitaevskii2004,Pitaevskii2011},  is negative~\cite{Pitaevskiisnake2008}.  This is a consequence of  the soliton energy decreasing  with increasing the travel speed.  Applying this equation to an element of a line soliton subject to long wavelength transverse perturbations:  the acceleration direction is opposite to the  restoring force, the amplitude of the transverse deformations  amplifies, and the soliton eventually breaks down into vortex dipoles~\cite{huang2003}. 
Recently, two types of ferrodark solitons (FDSs), manifesting as kinks in the transverse magnetization $F_{\perp}\equiv F_x+i F_y$ (with the magnetic field along   $z$), have been found in a ferromagnetic spin-1 BEC~\cite{MDWYuBlair, FDSexact2022, FDScore2022}.  For  type-I FDSs, the inertial mass is positive while for type-II FDSs it is negative~\cite{FDSexact2022}.   Hence, it could be expected that a type-II FDS in a quasi-two-dimensional (2D) system fragments under long wavelength transverse perturbations.  

In this Letter, however, we find that, instead of breaking down, type-II FDSs exhibit complex non-linear dynamics while preserving the  topological magnetic domain wall structure.  We study the dynamical instabilities and dynamics of FDSs in a quasi-2D spin-1 BEC. For type-II FDSs, two dynamical instabilities are found, being a snake instability and a spin-twist instability.  The unstable modes associated with the spin-twist instability generate spin-vortex dipoles in the $F_x-F_z$ plane. These spin-vortices are anomalous in that they do not have phase winding in the component wavefunctions. The amplitude of a long wavelength transverse perturbation to the type-II FDS grows at early times driven by the snake instability. 
This growth saturates at later time as elements reaching the propagating speed limit of FDSs convert to type-I FDSs. This leads to a hybridization of the soliton with spatially alternating  type-I and type-II segments along its length, and 
mass vortex dipoles in the $m=0$ component produced at nodes of the motion and confined by the line soliton. 
As a result, the hybridized FDS together with the $m=0$ mass vortex dipoles form a composite excitation and  exhibits non-periodic and complex 2D motion at long times without loosing the domain wall topological structure in the transverse magnetization.  In contrast, for the type-I FDSs, only the spin-twist instability exists and the combination of out-of-phase snake-like unstable modes in $m=\pm 1$ components  produces polar-core spin-vortex dipole pairs in the $F_x-F_y$ plane.  Since the inertial mass is positive, a type-I FDS oscillates under transverse perturbations, resembling the motion of an elastic string.   

\textit{Spin-1 BECs ---}
A spin-1 BEC  consists of atoms with three hyperfine states $\ket{F=1,m=+1,0,-1}$ and is described by  the three component wavefunction $\psi=(\psi_{+1},\psi_{0},\psi_{-1})^{T}$. The Hamiltonian density of a weekly interacting spin-1 BEC reads 
 \bea 
\label{Hamioltonian}
	{\cal H}=  \frac{\hbar^2 \left|\nabla \psi\right|^2 }{2M} +\frac{g_n}{2} |\psi^{\dag}\psi|^2+\frac{g_s}{2} |\psi^{\dag} \mathbf{S} \psi|^2 +q \psi^{\dag} S^2_z \psi,
\eea
where $M$ is the atomic mass, $g_n>0$ is the density interaction strength, $g_s$ is the spin-dependent interaction strength,
$\mathbf{S}=(S_x,S_y,S_z)$, and $S_{j=x,y,z}$ are the spin-1 matrices~\cite{StamperRMP,KAWAGUCHI12}. The uniform magnetic field is  along the $z$-axis, and  $q$ denotes the quadratic Zeeman energy.  The mean-field dynamics of the field $\psi$ is governed by the  Gross-Pitaevskii equations (GPEs) 
\begin{subequations}
\bea
\hspace{-2mm} i\hbar \frac{\partial \psi_{\pm 1}}{\partial t}&&=\left[H_0+g_s\left(n_0+n_{\pm 1}-n_{\mp 1}\right)+q \right]\psi_{\pm 1}+g_s \psi^2_0 \psi^{*}_{\mp 1},\,\, \\
i\hbar \frac{\partial \psi_0}{\partial t}&&= \left[H_0 +g_s\left(n_{+1}+n_{-1}\right) \right]\psi_0 + 2g_s \psi^{*}_0\psi_{+1}\psi_{-1},
\eea 
\label{spin-1GPE}
\end{subequations}
\hspace{-1mm}where $H_0=-\hbar^2\nabla^2/2M +g_n n$, $n_m=|\psi_m|^2$ and $n=\sum n_m$.  The last terms on the right-hand side of  Eqs.~\eqref{spin-1GPE} describe the internal coherent spin exchange dynamics: $\ket{00} \leftrightarrow\ket{+1}\ket{-1}$~\cite{Ho98,OM98,Stampernatrue2006}.
The magnetization density $\mathbf{F}\equiv\psi^{\dag} \mathbf{S} \psi$ is the order parameter quantifying  ferromagnetic  $|\mathbf{F}|>0$ for $g_s<0$ ($^{87}$Rb,$^7$Li)  and anti-ferromagnetic order $\mathbf{F}=0$ for $g_s>0$ ($^{23}$Na) \cite{Ho98,OM98,Stampernatrue2006,StamperRMP,KAWAGUCHI12}.   

\textit{FDSs---}
The easy-plane phase of a uniform ferromagnetic ($g_s<0$) spin-1 BEC is characterized by a transverse magnetization  $\mathbf{F}\to F_{\perp}=F^g_{\perp}(\tau)=\sqrt{8n^b_{\pm1}n^{b}_0}e^{i\tau}$, where $n^b_{\pm 1}=(1-\tilde{q})n_b/4$ and $n^b_0=n_b(1+\tilde{q})/2$ are the component densities,  $n_b$ is the total number density, and $\tilde{q}\equiv-q/(2g_sn_b)$. Under constraint $F_z=n_{+1}-n_{-1}=0$, this phase is the ground state for $0<\tilde{q}<1$~\cite{StamperRMP,KAWAGUCHI12}. The $\textrm{SO}(2)$ symmetry of the easy-plane state is parameterized by the rotational angle about the $z$-axis $\tau$. 
The FDS is a $\mathbb{Z}_2$ topological defect  in the magnetic order of the easy-plane phase and connects regions transversely magnetized in opposite directions ($\tau=0$ and $\tau=\pi$)~\cite{MDWYuBlair, FDSexact2022, FDScore2022}.  In 1D, it is an Ising type kink and the corresponding magnetization  $F_{\perp}(x)$ satisfies 
$F_{\perp}(-\infty)=F^{g}_{\perp}(\pi) \,\, \text{and} \,\, F_{\perp}(+\infty)=F^{g}_{\perp}(0)=-F_{\perp}(-\infty)$.  For $g_s=-g_n/2$ and $0< \tilde{q}<1$,  exact FDS solutions to Eqs.~\eqref{spin-1GPE} are available~\cite{FDSexact2022}. 
There exists two types of FDSs:  type-I FDSs have positive inertial mass while  type-II FDSs have negative inertial mass.  At the maximum speed,  the two types of FDSs  become identical and transitions between them could occur,  e.g., allowing oscillations (in space and type) of a FDS in hard-wall confined quasi-1D spin-1 BEC subject to a linear potential~\cite{FDSexact2022}. 
As we will see later, this property  plays an essential role in preventing fragmentation of a  line type-II FDS in its decaying dynamics driven by a dynamically unstable mode. 

\textit{Dynamical instabilities---}
The exact wavefunctions of the stationary FDSs read~\cite{FDSexact2022,FDScore2022}
\bea
\psi^{\rm I}_s&=&\left[\sqrt{n^g_{\pm 1}} \tanh \left(\frac{x}{2 \ell^{\rm I}}\right),\sqrt{n^g_0},\sqrt{n^g_{\pm 1}} \tanh \left(\frac{x}{2 \ell^{\rm I}}\right) \right]^{T}, \\
\psi^{\rm II}_s&=&\left[ \sqrt{n^g_{\pm 1}} , \sqrt{n^g_0} \tanh\left(\frac{x}{2\ell^{\rm II}}\right),\sqrt{n^g_{\pm 1}} \, \right]^{T}, 
\eea
where $\ell^{\rm I,II}=\hbar/\!\sqrt{2g_n n_b M (1\mp\tilde{q})}$, and minus and plus signs in front of $\tilde{q}$ are referred to as type-I and type-II FDSs, respectively. The corresponding transverse magnetization and  total number density are
$F^{\rm I,II}_{\perp}(x)=F^{\rm I,II}_{x}(x)=|F_{\perp}^g| \tanh(x/2\ell^{\rm I,II})$ and 
$n^{\rm I,II}(x)=n_b[1-(1\mp \tilde{q})/2\sech^2(x/2\ell^{\rm I,II})]$.
Away from the exactly solvable region, $\psi^{\rm I,II}_s$ can be obtained numerically~\cite{FDScore2022}. 

Let us now consider a straight infinitely long FDS along the $y$-axis located in the middle of a slab with width much larger than $\ell^{\rm I, \rm II}$.
We consider a perturbed FDS $\psi=\psi_s+\delta \psi$, where $\psi_s$ denotes a  stationary FDS and $\delta \psi=\epsilon[u(\mathbf{r})e^{-i\omega t}+v^{*}(\mathbf{r})e^{i \omega^{*} t}]$ is the perturbation with the dimensionless control parameter $\epsilon \ll 1$.   Keeping leading order terms in Eq.~\eqref{spin-1GPE} we obtain the Bogoliubov-de Gennes equations (BdGs): 
\bea
\hbar \omega  \left( {\begin{array}{cc}
		u\\
		v\\
\end{array} } \right)=
\left( {\begin{array}{cc}
		{\cal L}_{\rm GP}+X-\mu & \Delta \\
		-\Delta^{*} & -({\cal L}_{\rm GP}+X-\mu)^{*} \\
\end{array} } \right)
\left( {\begin{array}{cc}
		u\\
		v\\
\end{array} } \right),
\label{BdG1}
\eea
where the stationary wavefunction satisfies  ${\cal L}_{\rm GP}\psi_s=\mu \psi_s$, $X= g_s\sum_{j=x,y,z} S_{j}\psi_s\psi_s^{\dagger}S_j+ g_n \psi_{s}\psi^{\dagger}_s $, $\Delta=g_n \psi_s \psi^{T}_s+g_s \sum_{j=x,y,z} (S_j \psi_s)(S_j\psi_s)^{T}$ and $\mu=(g_n+g_s)n_b+q/2$ is the chemical potential.  Since the system has translational symmetry along the $y-$axis, we can use the wavevector $k_y$ to parameterize the perturbations as $u(\mathbf{r})=u(x) e^{ik_y y}$ and $v(\mathbf{r})=v(x) e^{ik_y y}$.  An imaginary part of energy $\hbar \omega$ marks a  dynamical instability in the system, i.e., a mode that grows exponentially with time.  

We numerically solve Eq.~\eqref{BdG1} with Neumann boundary conditions at $x$-axis boundaries~\cite{footnoteunstable}. 
\begin{figure}[htp] 
	\centering
	\includegraphics[width=0.418\textwidth]{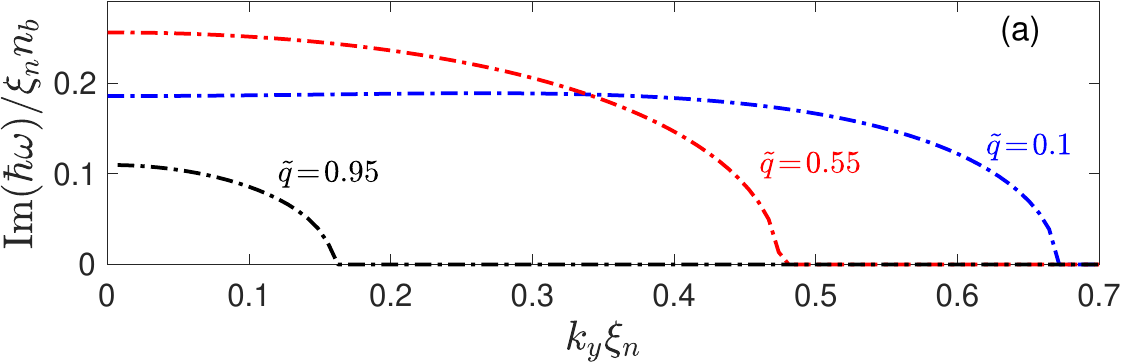}
	\includegraphics[width=0.418\textwidth]{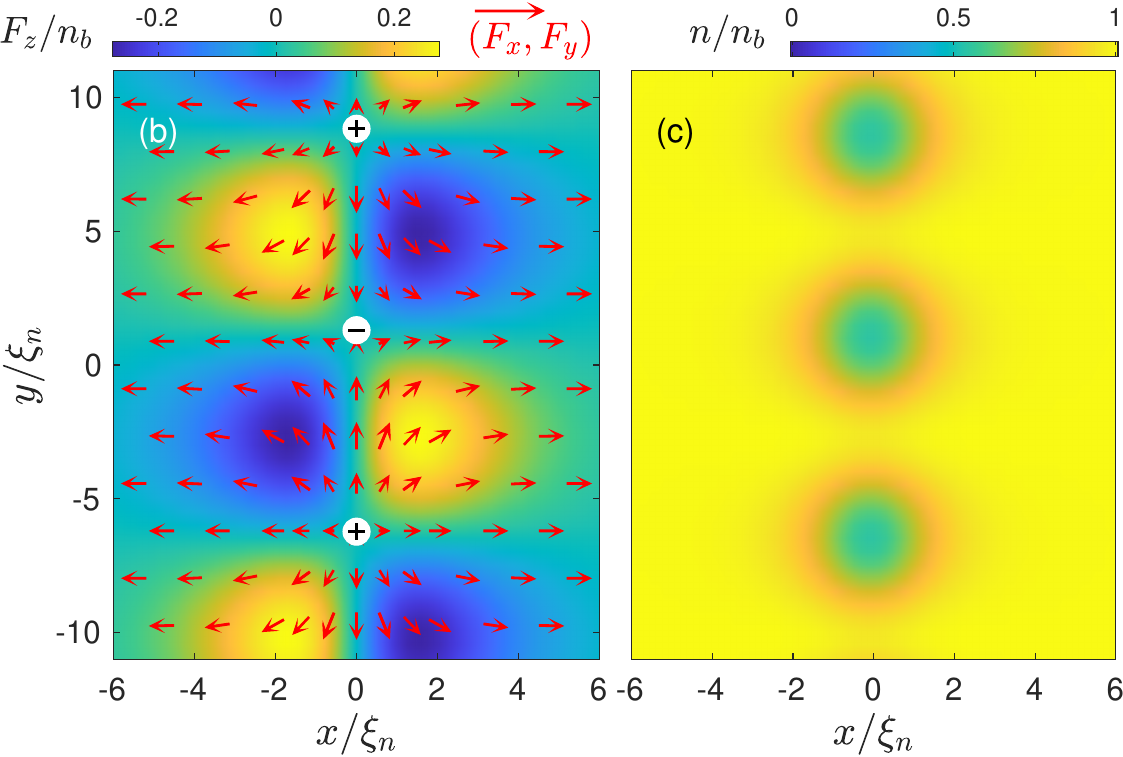}
	\includegraphics[width=0.418\textwidth]{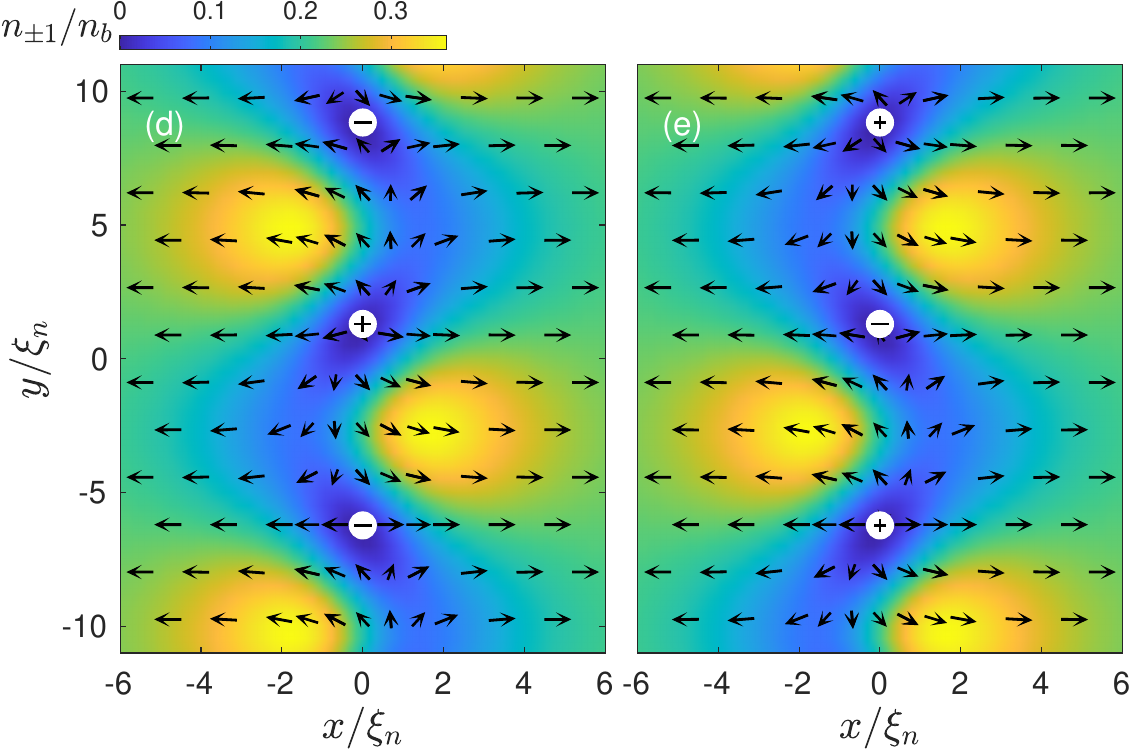}
	\caption{Unstable modes associated with the spin-twist instability of type-I FDSs. (a)  Imaginary of part of the spectrum of the unstable modes for $g_s/g_n=-1/2$ and $\tilde{q}=0.1, 0.55,0.95$. For the unstable mode [$\rm{Im} (\hbar \omega) \neq0$], the corresponding  real part of the spectrum is zero.  (b) Spin-texture created by the unstable mode.  White circles with $+$ and $-$ indicate positive and negative circulation polar-core spin-vortices, respectively. The red arrows and the background color represent the transverse magnetization field $(F_x , F_y )$ and longitudinal magnetization $F_z$, respectively.  Moving along $y$ (at constant $x$) $F_y$ modulates while $F_x$ is constant.  Here $n_b$ is the ground state density and $\xi_n=\hbar/\sqrt{M g_n n_b}$ is the density healing length. (c) Total density modulation created by the unstable modes. In (d) and (e), the background color represent the profile of $n_{+1}$ and $n_{-1}$ and the black arrows represent the phase of $\psi_{+1}$ and $\psi_{-1}$, respectively.  White circles  indicate   vortices in the components. The unstable modes in (b)-(e) are chosen at $k_y \xi_n \simeq 0.417$, $\tilde{q}=0.1$ and we use  $\epsilon=2$ for visibility purposes.   } 
	\label{f:unstablemodestypeI}
\end{figure} 
For type-I FDSs,  there is only one dynamical instability and the $k_y$ region associated with the imaginary energy decreases as $\tilde{q}$  increases [Fig.~\ref{f:unstablemodestypeI}(a)]. The unstable mode leaves $F_x$ unchanged and generates polar-core spin vortex dipole pairs in the $(F_x,F_y)$ plane combined with modulations of $F_z$ [Fig.~\ref{f:unstablemodestypeI}(b)] and  $n$  [Fig.~\ref{f:unstablemodestypeI}(c)].   We refer to this instability as type-I spin-twist instability.  Interestingly, we see that this arises from the \textit{out-of-phase combination} of snake-like unstable modes in $m=\pm 1$  components [Figs.~\ref{f:unstablemodestypeI}(d) and \ref{f:unstablemodestypeI}(e)].   

\begin{figure}[htp] 
	\centering
	\includegraphics[width=0.418\textwidth]{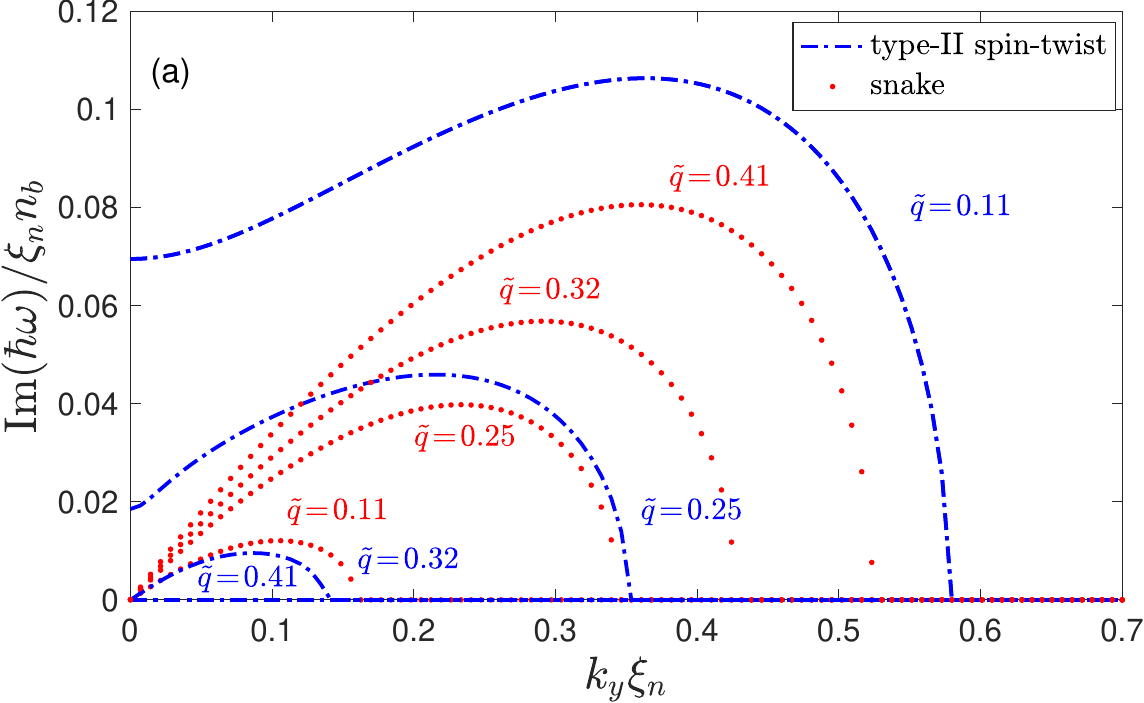}
	\includegraphics[width=0.418\textwidth]{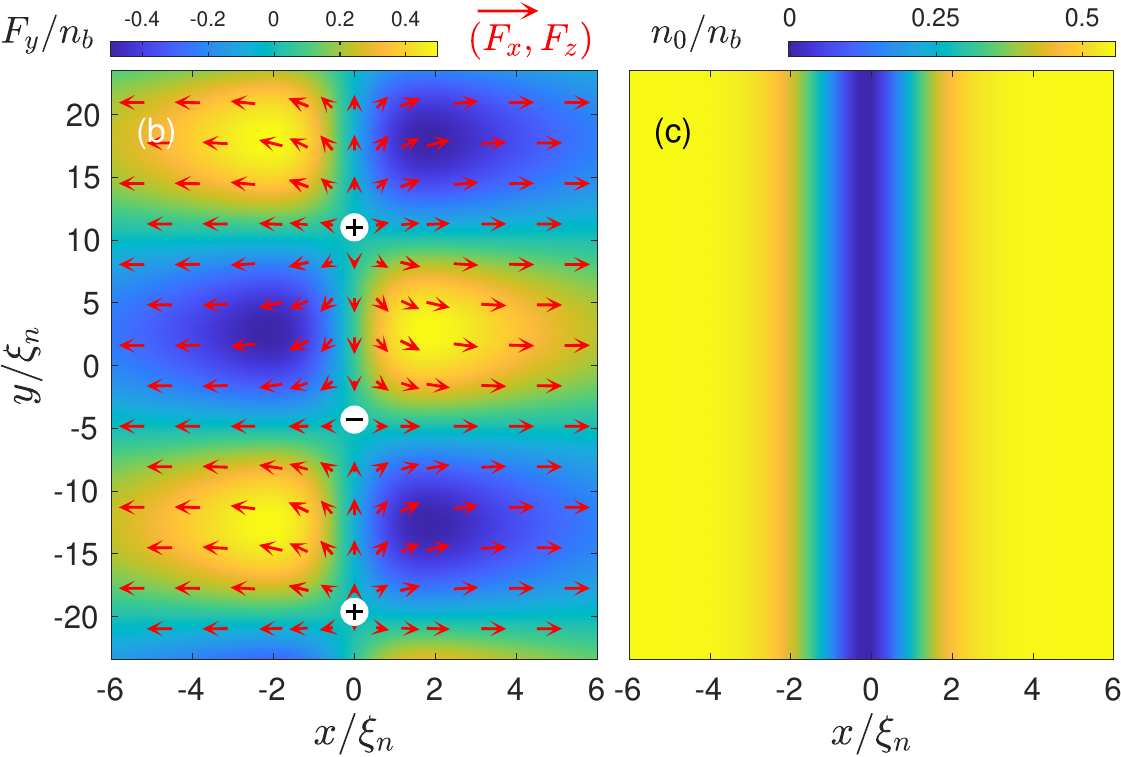}
	\includegraphics[width=0.418\textwidth]{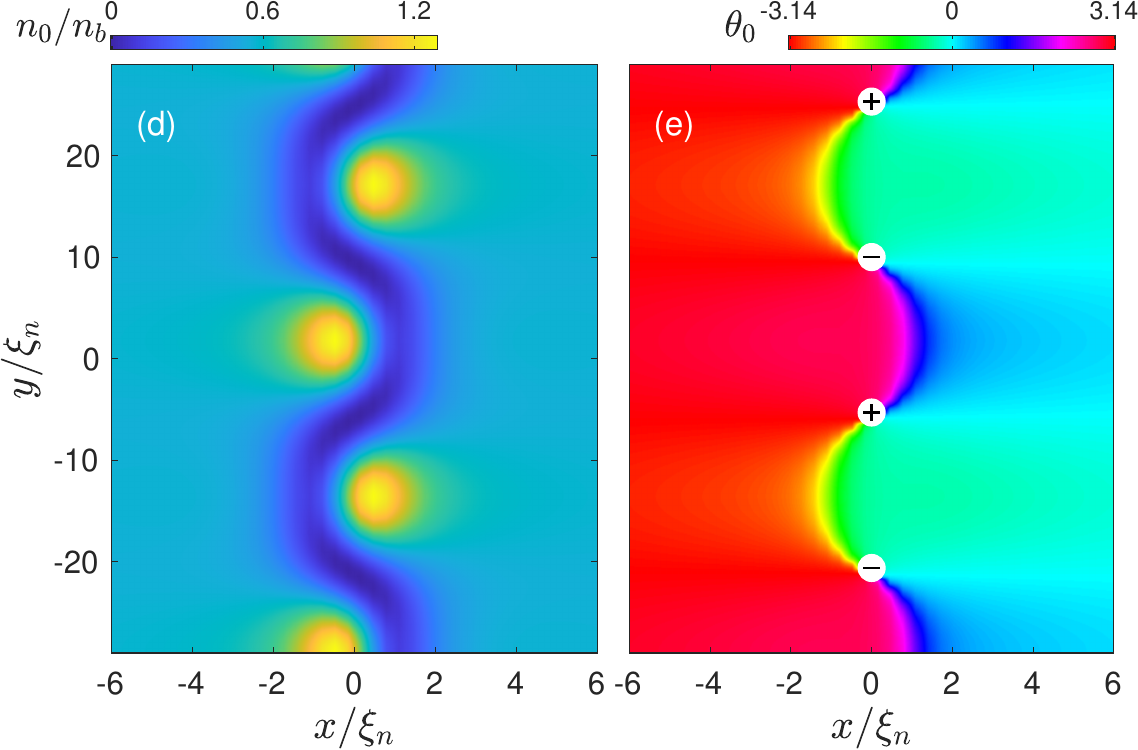}
	\caption{Two unstable modes associated with type-II FDS 
	(a)  Imaginary of part of the spectrum of the unstable modes for $g_s/g_n=-1/2$ and $\tilde{q}=0.11, 0.25, 0.32, 0.41$ (corresponding  real part of the spectrum is zero).  (b), (c) Effect of spin-twist unstable mode on the FDS. (b)  White circles with $+$ and $-$ indicate positive and negative circulation of $(F_x , F_z )$ spin-vortices, respectively. The red arrows and the background color represent magnetization fields $(F_x , F_z )$ and  $F_y$, respectively.   (c) Component density $n_0$ is unaffected. (d), (e) Effect of snake unstable mode on the FDS.  (d) Component density $n_0$. (e) Phase $\theta_0$  of the $m=0$ component, with white circles  indicating the circulation of $\psi_0$ component vortices. Accordingly,  $F_x$ is only spatially deformed and  $n_{\pm1}$ is weakly modulated (see Figs. S1(f) and S1(g)~\cite{SM}). The unstable modes  in (b)-(e) are chosen  at $k_y \xi_n \simeq 0.205$,  $\tilde{q}=0.25$ and we  use  $\epsilon=2$ for visibility purposes. } 
	\label{f:instabilitytypeII}
\end{figure}

For type-II FDSs, there exist two distinct unstable modes  [Fig.~\ref{f:instabilitytypeII}(a)].   One unstable mode, which we refer to as the type-II spin-twist instability,  creates spin vortex dipole pairs in the $(F_x,F_z)$ plane accompanied by the modulations of $F_y$, while $F_x$ is unchanged~\cite{spintwist}.  Noticeably, in contrast to conventional spin vortices (polar core and Mermin-Ho vortices)~\cite{KAWAGUCHI12},  these spin-vortices are not generated by phase winding in components [Fig.~\ref{f:instabilitytypeII}(b)]. 
For given $g_s/g_n$, the unstable $k_y$ region decreases monotonically as $\tilde{q}$ increases and the imaginary energy vanishes at a certain value of $\tilde{q}$ [Fig.~\ref{f:instabilitytypeII}(a)]~\cite{footnotesmallgs}. 
The second unstable mode is a snake instability arising from the negative inertial mass of the type-II FDS.  The imaginary energy as a function of $k_y$ for this instability [see Fig.~\ref{f:instabilitytypeII}(a)]  is similar to that for a dark soliton in scalar BECs~\cite{Shlyapnikov1999}. For given $g_s/g_n$, the snake instability grows as  $\tilde{q}$ increases, since the distinction between two types of FDSs is greater for larger $\tilde{q}$ [Fig.~\ref{f:instabilitytypeII}(a)]. 
The unstable mode resides dominantly in $m=0$ state and creates transverse deformation of component density $n_0$ and vortex-dipole pairs  [Figs.~\ref{f:instabilitytypeII}(d) and \ref{f:instabilitytypeII} (c)],  a typical feature of the snake instability~\cite{footzeroq}.  

\begin{figure}[htp] 
	\centering
	\includegraphics[width=0.418\textwidth]{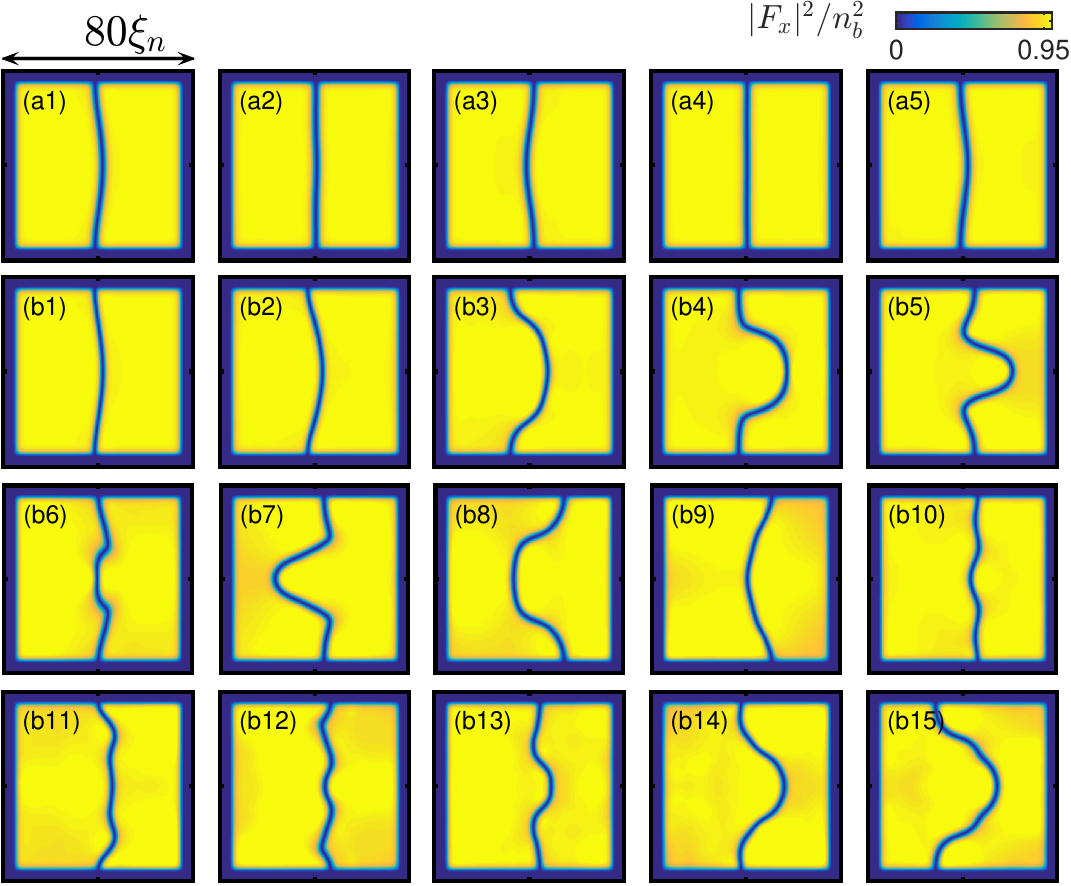}
	\caption{ Comparison between dynamics of  open type-I and type-II FDSs with free end-points attached on the hard-wall boundaries,  starting from initial states under the same transverse deformation  [(a1) and (b1)].  The  parameters  are  $g_s/g_n=-1/2$ and $\tilde{q}=0.25$. The system is  confined by hard-wall potentials in a square domain $x, y \in [-L, L]$ with $L = 40\xi_n$.  The deformed profile is described by a cosine function. The time interval between snapshots is $60 t_0$, where $t_0=\hbar/g_n  n_b$. (a1-a5) One period of evolution of the transversely deformed  type-I FDS.  (b1-b5)  Evolution of a type-II FDS  at the same times as in (a). (b6-b15) Subsequent  time evolution  after (b5).  Subsequent evolution after (b15) for much longer times is shown in Fig. S2~\cite{SM}. } 
	\label{f:oscillation}
\end{figure}

\textit{Absence of breakdown of type-II FDSs and proliferation of confined vortex dipoles---}  Let us now consider an open FDS in a spin-1 BEC  confined by a hard-wall potential in a 2D square domain and study its real time dynamics. We initially impose a cosinusoidal transverse perturbation, with a  system-size wavelength  (much larger than the soliton width),   to the FDS [Fig.~\ref{f:oscillation} a(1)].  
For a type-I FDS, since the inertial mass is positive,  it exhibits elastic oscillations [Figs.~\ref{f:oscillation}(a1)-\ref{f:oscillation}(a5)]~\cite{footnotepositivemass}.  Applying the same transverse deformation to  the type-II FDS  (Fig.~\ref{f:oscillation} [b(1)]), its dynamics are rather different:  the magnitude of the transverse perturbation increases initially with time [Figs.~\ref{f:oscillation}(b1)-\ref{f:oscillation}(b3)], as an unavoidable consequence of the negative inertial mass and the snake instability [Figs.~\ref{f:instabilitytypeII}(d) and \ref{f:instabilitytypeII}(e)].
We would normally expect that the type-II FDS breaks down at later times.
However the initial linear growth of the amplitude saturates and  subsequently the FDS enters a nonlinear regime. Significantly,  the FDS topological structure persists without fragmenting into pieces and exhibits non-periodic and complex nonlinear dynamics at late times [Figs.~\ref{f:oscillation}(b4)-\ref{f:oscillation}(b15)].  
\begin{figure}[htp] 
	\centering
	\hspace{0cm}
	\includegraphics[trim = 0mm 0mm 0mm 0mm, clip, width=0.418\textwidth]{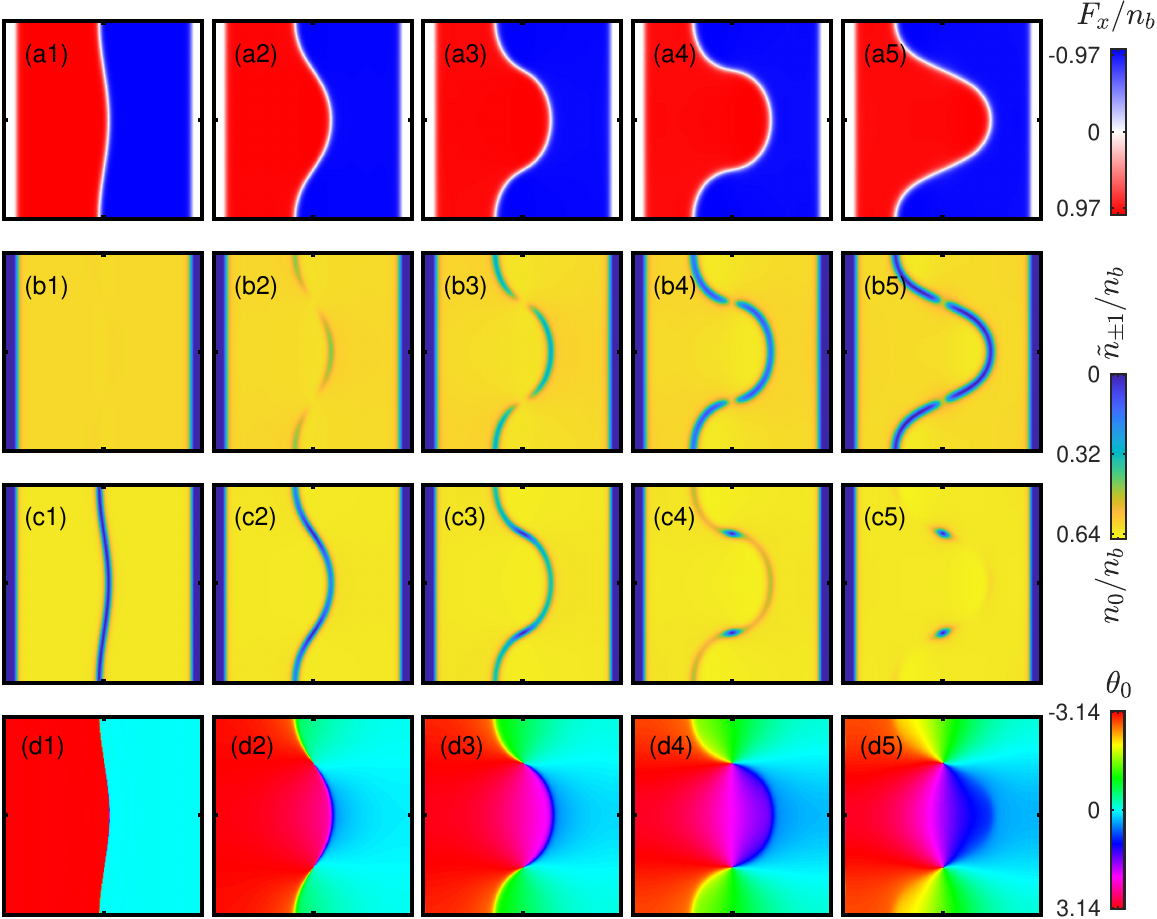}
	\vfill
	\hspace{-1.3cm}
	\includegraphics[trim = 0mm 0mm 0mm 0mm, clip, width=0.408\textwidth]{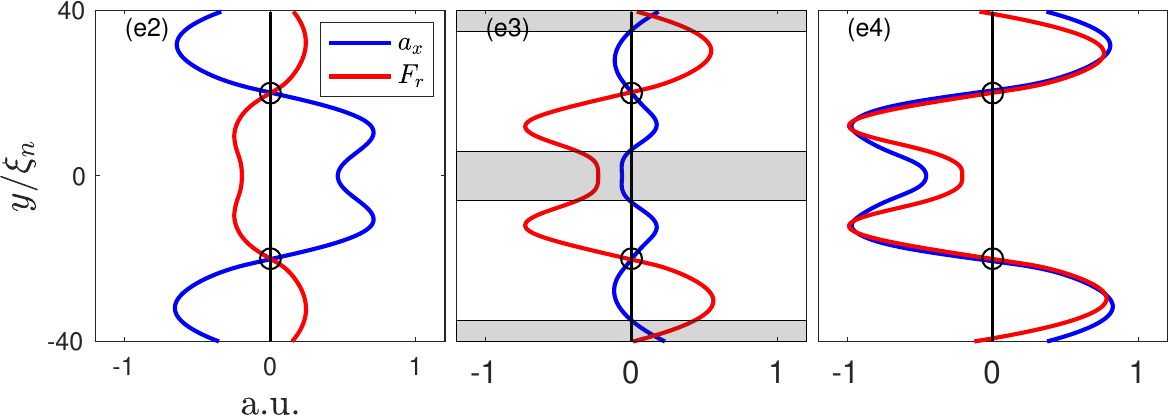}
	\caption{Early time dynamics of a type-II FDS.  The system is periodic  in $y$ and is confined by hard-wall potentials in $x$.  The time interval between snapshots is $30 t_0$. The initial transverse deformation and other parameters are as in Fig.~\ref{f:oscillation}.   (a1)-(a5) Profile of the transverse magnetization $F_x$.  The sign change of $F_{x}$ across the core  is kept during the motion.   (b1)-(b5) and (c1)-(c5) dynamics of component densities $n_{\pm 1}$ and $n_0$, respectively.  Here  $\tilde{n}_{\pm 1}=n_{\pm} \bar{n}_0/\bar{n}_{\pm1}$ is re-scaled component density and $\bar{n}_0$ and $\bar{n}_{\pm 1}$ are ground state component densities.  (d1)-(d5) The phase of $m=0$ spin state, showing proliferation of a mass vortex dipole in $m=0$ component [(d4),(d5)]. (e2)-(e4) Acceleration (blue line) and restoring force (red line) at evolution stages (2), (3) and (4). In (e3), $F_r$ and $a_x$ have the same sign in the shaded region.  The values  of $F_r$ and $a_x$ are in arbitrary units and are scaled within $[-1,1]$ for convenience. The black circles mark the positions of nodes [$(0,-L/4),(0,L/4)$] which coincide with the positions of $m=0$ vortices.}
	\label{f:typeIIphase}
\end{figure} 

In order to examine the process of entering the nonlinear regime for an unstable type-II FDS, we consider a system that is periodic in the $y$ direction (Fig.~\ref{f:typeIIphase}) for simplicity (to avoid the influence of boundaries).
Let $x=x(y,t)$ be the position of the soliton core. The restoring force $F_r(y,t) =T \partial^2 x/\partial y^2$ and the acceleration $a_x(y,t)=\partial^2 x/\partial t^2$ have same sign for type-I FDSs and have opposite signs for type-II FDSs, where $T>0$ is the ``string'' tension. At early times, the soliton is still a type-II FDS, and $F_r(y,t)$ and $a_x(y,t)$  have opposite signs [Fig.~\ref{f:typeIIphase}(e2)].  When a section of the soliton reaches the maximum speed, and it transfers into a type-I FDS locally, signaled by the corresponding part of  $F_r(y,t)$ and $a_x(y,t)$ having same sign  [Fig.~\ref{f:typeIIphase}(e3)].  At this stage the soliton forms a hybridized chain, with alternating line segments of type-I and type-II FDS character.  At later times in the decaying dynamics,  the line type-II FDS transfers entirely to a type-I FDS [Fig.~\ref{f:typeIIphase}(e4)]~\cite{typetransitions}. Additionally,  ``virtual''  $m=0$ mass  vortex-dipole pairs which are encoded in the unstable modes associated with the  snake instability [see Fig.~\ref{f:instabilitytypeII}(e)]  become real vortex dipole pairs at the places where the soliton speed is zero [Figs.~\ref{f:typeIIphase}(c1)-\ref{f:typeIIphase}(c5)].  Intriguingly,  these $m=0$ mass vortex dipoles are confined by the type-I FDS and do not become free.  Hence the topological structure of the magnetic domain wall is preserved and a stable vortex-domain wall composite excitation forms as  a preferred path  of the decaying dynamics.  The interplay between the mass current generated by the confined $m=0$ vortex dipoles and a hybridized FDS  gives rise to rather complex and rich  nonlinear dynamics of the line composite excitation [Figs.~\ref{f:oscillation}(b6)-\ref{f:oscillation}(b15)].  Note that the $ m=0$ mass vortices are $\mathbb{U}(1)$ topological defects which are responsible for breaking the mass superfluid order~\cite{Andrew2023} and the FDSs are  $\mathbb{Z}_2$ topological defects in the spin order.  Hence the two defects can coexist and form a composite excitation, revealing a deep relation between the two orders of the spinor superfluid in the nonlinear regime. 

\textit{Conclusion---}  
We investigate dynamical instability and dynamics of line  ferrodark solitons  in a quasi-2D ferromagnetic spin-1 Bose-Einstein condensate.  
A type-II ferrodark soliton processes both snake  and spin-twist instabilities. 
In contrast to the normal situation, however,  in dynamics a type-II ferrodark soliton does not fragment under transverse perturbations and instead exhibits complex motion accompanied by forming a hybridized chain of type-I and type-II  ferrodark soliton with a proliferation of confined $m=0$ vortex dipole pairs.  
A type-I ferrodark soliton exhibits oscillations under transverse perturbations~\cite{footnotefrequency} and the relevant unstable mode creates polar-core spin-vortex dipoles. This work explores rich phenomena of 2D soliton instability and dynamics~\cite{footnotegeneral} which have not been encountered in the past and will motivate further theoretical studies.  
Moreover, due to the robustness of line  ferrodark solitons against their dynamical instabilities,  they may play an important role in finite temperature phase transitions in 2D spinor superfluids~\cite{Lamacraft2011}.  Also, the stable vortex-ferrodark composite exhibits rich dynamics of the $m=0$ vortex dipoles  confined by the soliton,  offering another opportunity to simulate some aspects of  the confinement physics in ultra cold gas systems~\cite{Son2002, Tylutki2016}. Furthermore, experimental investigations are also within the scope of current ultracold-gas experiments~\cite{Dalibard2015,Gauthier16,Semeghini2018,Higbie2005,Huh2020a,MSexp1, MSexp2, prufer2022condensation}.

\textit{Acknowledgment---}
We thank M. Antonio and  D. Baillie for useful discussions. X.Y. acknowledges support from the National Natural Science Foundation of China (Grant No. 12175215, Grant No. 12475041), the National Key Research and Development Program of China (Grant No. 2022YFA 1405300) and  NSAF (Grant No. U2330401). P.B.B acknowledges support from the Marsden Fund of the Royal Society of New Zealand.


\begin{thebibliography}{40}%
	\makeatletter
	\providecommand \@ifxundefined [1]{%
		\@ifx{#1\undefined}
	}%
	\providecommand \@ifnum [1]{%
		\ifnum #1\expandafter \@firstoftwo
		\else \expandafter \@secondoftwo
		\fi
	}%
	\providecommand \@ifx [1]{%
		\ifx #1\expandafter \@firstoftwo
		\else \expandafter \@secondoftwo
		\fi
	}%
	\providecommand \natexlab [1]{#1}%
	\providecommand \enquote  [1]{``#1''}%
	\providecommand \bibnamefont  [1]{#1}%
	\providecommand \bibfnamefont [1]{#1}%
	\providecommand \citenamefont [1]{#1}%
	\providecommand \href@noop [0]{\@secondoftwo}%
	\providecommand \href [0]{\begingroup \@sanitize@url \@href}%
	\providecommand \@href[1]{\@@startlink{#1}\@@href}%
	\providecommand \@@href[1]{\endgroup#1\@@endlink}%
	\providecommand \@sanitize@url [0]{\catcode `\\12\catcode `\$12\catcode
		`\&12\catcode `\#12\catcode `\^12\catcode `\_12\catcode `\%12\relax}%
	\providecommand \@@startlink[1]{}%
	\providecommand \@@endlink[0]{}%
	\providecommand \url  [0]{\begingroup\@sanitize@url \@url }%
	\providecommand \@url [1]{\endgroup\@href {#1}{\urlprefix }}%
	\providecommand \urlprefix  [0]{URL }%
	\providecommand \Eprint [0]{\href }%
	\providecommand \doibase [0]{http://dx.doi.org/}%
	\providecommand \selectlanguage [0]{\@gobble}%
	\providecommand \bibinfo  [0]{\@secondoftwo}%
	\providecommand \bibfield  [0]{\@secondoftwo}%
	\providecommand \translation [1]{[#1]}%
	\providecommand \BibitemOpen [0]{}%
	\providecommand \bibitemStop [0]{}%
	\providecommand \bibitemNoStop [0]{.\EOS\space}%
	\providecommand \EOS [0]{\spacefactor3000\relax}%
	\providecommand \BibitemShut  [1]{\csname bibitem#1\endcsname}%
	\let\auto@bib@innerbib\@empty
	\bibitem [{\citenamefont {Kuznetsov}\ and\ \citenamefont
		{Turitsyn}(1988)}]{kuznetsov1988instability}%
	\BibitemOpen
	\bibfield  {author} {\bibinfo {author} {\bibfnamefont {E.}~\bibnamefont
			{Kuznetsov}}\ and\ \bibinfo {author} {\bibfnamefont {S.}~\bibnamefont
			{Turitsyn}},\ }\href@noop {} {\bibfield  {journal} {\bibinfo  {journal} {Zh.
				Eksp. Teor. Fiz}\ }\textbf {\bibinfo {volume} {94}},\ \bibinfo {pages} {129}
		(\bibinfo {year} {1988})}\BibitemShut {NoStop}%
	\bibitem [{\citenamefont {Muryshev}\ \emph {et~al.}(1999)\citenamefont
		{Muryshev}, \citenamefont {van Linden van~den Heuvell},\ and\ \citenamefont
		{Shlyapnikov}}]{Shlyapnikov1999}%
	\BibitemOpen
	\bibfield  {author} {\bibinfo {author} {\bibfnamefont {A.~E.}\ \bibnamefont
			{Muryshev}}, \bibinfo {author} {\bibfnamefont {H.~B.}\ \bibnamefont {van
				Linden van~den Heuvell}}, \ and\ \bibinfo {author} {\bibfnamefont {G.~V.}\
			\bibnamefont {Shlyapnikov}},\ }\href {\doibase 10.1103/PhysRevA.60.R2665}
	{\bibfield  {journal} {\bibinfo  {journal} {Phys. Rev. A}\ }\textbf {\bibinfo
			{volume} {60}},\ \bibinfo {pages} {R2665} (\bibinfo {year}
		{1999})}\BibitemShut {NoStop}%
	\bibitem [{\citenamefont {Scott}\ \emph {et~al.}(2011)\citenamefont {Scott},
		\citenamefont {Dalfovo}, \citenamefont {Pitaevskii},\ and\ \citenamefont
		{Stringari}}]{Pitaevskii2011}%
	\BibitemOpen
	\bibfield  {author} {\bibinfo {author} {\bibfnamefont {R.~G.}\ \bibnamefont
			{Scott}}, \bibinfo {author} {\bibfnamefont {F.}~\bibnamefont {Dalfovo}},
		\bibinfo {author} {\bibfnamefont {L.~P.}\ \bibnamefont {Pitaevskii}}, \ and\
		\bibinfo {author} {\bibfnamefont {S.}~\bibnamefont {Stringari}},\ }\href
	{\doibase 10.1103/PhysRevLett.106.185301} {\bibfield  {journal} {\bibinfo
			{journal} {Phys. Rev. Lett.}\ }\textbf {\bibinfo {volume} {106}},\ \bibinfo
		{pages} {185301} (\bibinfo {year} {2011})}\BibitemShut {NoStop}%
	\bibitem [{\citenamefont {Gallem\'{\i}}\ \emph {et~al.}(2019)\citenamefont
		{Gallem\'{\i}}, \citenamefont {Pitaevskii}, \citenamefont {Stringari},\ and\
		\citenamefont {Recati}}]{Gallemi2019}%
	\BibitemOpen
	\bibfield  {author} {\bibinfo {author} {\bibfnamefont {A.}~\bibnamefont
			{Gallem\'{\i}}}, \bibinfo {author} {\bibfnamefont {L.~P.}\ \bibnamefont
			{Pitaevskii}}, \bibinfo {author} {\bibfnamefont {S.}~\bibnamefont
			{Stringari}}, \ and\ \bibinfo {author} {\bibfnamefont {A.}~\bibnamefont
			{Recati}},\ }\href {\doibase 10.1103/PhysRevA.100.023607} {\bibfield
		{journal} {\bibinfo  {journal} {Phys. Rev. A}\ }\textbf {\bibinfo {volume}
			{100}},\ \bibinfo {pages} {023607} (\bibinfo {year} {2019})}\BibitemShut
	{NoStop}%
	\bibitem [{\citenamefont {Qu}\ \emph {et~al.}(2017)\citenamefont {Qu},
		\citenamefont {Tylutki}, \citenamefont {Stringari},\ and\ \citenamefont
		{Pitaevskii}}]{MDQu2017}%
	\BibitemOpen
	\bibfield  {author} {\bibinfo {author} {\bibfnamefont {C.}~\bibnamefont
			{Qu}}, \bibinfo {author} {\bibfnamefont {M.}~\bibnamefont {Tylutki}},
		\bibinfo {author} {\bibfnamefont {S.}~\bibnamefont {Stringari}}, \ and\
		\bibinfo {author} {\bibfnamefont {L.~P.}\ \bibnamefont {Pitaevskii}},\ }\href
	{\doibase 10.1103/PhysRevA.95.033614} {\bibfield  {journal} {\bibinfo
			{journal} {Phys. Rev. A}\ }\textbf {\bibinfo {volume} {95}},\ \bibinfo
		{pages} {033614} (\bibinfo {year} {2017})}\BibitemShut {NoStop}%
	\bibitem [{\citenamefont {Shamailov}\ and\ \citenamefont
		{Brand}(2018)}]{Sophie}%
	\BibitemOpen
	\bibfield  {author} {\bibinfo {author} {\bibfnamefont {S.~S.}\ \bibnamefont
			{Shamailov}}\ and\ \bibinfo {author} {\bibfnamefont {J.}~\bibnamefont
			{Brand}},\ }\href {\doibase 10.21468/SciPostPhys.4.3.018} {\bibfield
		{journal} {\bibinfo  {journal} {SciPost Phys.}\ }\textbf {\bibinfo {volume}
			{4}},\ \bibinfo {pages} {18} (\bibinfo {year} {2018})}\BibitemShut {NoStop}%
	\bibitem [{\citenamefont {Qu}\ \emph {et~al.}(2016)\citenamefont {Qu},
		\citenamefont {Pitaevskii},\ and\ \citenamefont {Stringari}}]{MDQu2016}%
	\BibitemOpen
	\bibfield  {author} {\bibinfo {author} {\bibfnamefont {C.}~\bibnamefont
			{Qu}}, \bibinfo {author} {\bibfnamefont {L.~P.}\ \bibnamefont {Pitaevskii}},
		\ and\ \bibinfo {author} {\bibfnamefont {S.}~\bibnamefont {Stringari}},\
	}\href {\doibase 10.1103/PhysRevLett.116.160402} {\bibfield  {journal}
		{\bibinfo  {journal} {Phys. Rev. Lett.}\ }\textbf {\bibinfo {volume} {116}},\
		\bibinfo {pages} {160402} (\bibinfo {year} {2016})}\BibitemShut {NoStop}%
	\bibitem [{\citenamefont {Farolfi}\ \emph {et~al.}(2020)\citenamefont
		{Farolfi}, \citenamefont {Trypogeorgos}, \citenamefont {Mordini},
		\citenamefont {Lamporesi},\ and\ \citenamefont {Ferrari}}]{MSexp1}%
	\BibitemOpen
	\bibfield  {author} {\bibinfo {author} {\bibfnamefont {A.}~\bibnamefont
			{Farolfi}}, \bibinfo {author} {\bibfnamefont {D.}~\bibnamefont
			{Trypogeorgos}}, \bibinfo {author} {\bibfnamefont {C.}~\bibnamefont
			{Mordini}}, \bibinfo {author} {\bibfnamefont {G.}~\bibnamefont {Lamporesi}},
		\ and\ \bibinfo {author} {\bibfnamefont {G.}~\bibnamefont {Ferrari}},\ }\href
	{\doibase 10.1103/PhysRevLett.125.030401} {\bibfield  {journal} {\bibinfo
			{journal} {Phys. Rev. Lett.}\ }\textbf {\bibinfo {volume} {125}},\ \bibinfo
		{pages} {030401} (\bibinfo {year} {2020})}\BibitemShut {NoStop}%
	\bibitem [{\citenamefont {Chai}\ \emph {et~al.}(2020)\citenamefont {Chai},
		\citenamefont {Lao}, \citenamefont {Fujimoto}, \citenamefont {Hamazaki},
		\citenamefont {Ueda},\ and\ \citenamefont {Raman}}]{MSexp2}%
	\BibitemOpen
	\bibfield  {author} {\bibinfo {author} {\bibfnamefont {X.}~\bibnamefont
			{Chai}}, \bibinfo {author} {\bibfnamefont {D.}~\bibnamefont {Lao}}, \bibinfo
		{author} {\bibfnamefont {K.}~\bibnamefont {Fujimoto}}, \bibinfo {author}
		{\bibfnamefont {R.}~\bibnamefont {Hamazaki}}, \bibinfo {author}
		{\bibfnamefont {M.}~\bibnamefont {Ueda}}, \ and\ \bibinfo {author}
		{\bibfnamefont {C.}~\bibnamefont {Raman}},\ }\href {\doibase
		10.1103/PhysRevLett.125.030402} {\bibfield  {journal} {\bibinfo  {journal}
			{Phys. Rev. Lett.}\ }\textbf {\bibinfo {volume} {125}},\ \bibinfo {pages}
		{030402} (\bibinfo {year} {2020})}\BibitemShut {NoStop}%
	\bibitem [{\citenamefont {Konotop}\ and\ \citenamefont
		{Pitaevskii}(2004)}]{Pitaevskii2004}%
	\BibitemOpen
	\bibfield  {author} {\bibinfo {author} {\bibfnamefont {V.~V.}\ \bibnamefont
			{Konotop}}\ and\ \bibinfo {author} {\bibfnamefont {L.}~\bibnamefont
			{Pitaevskii}},\ }\href {\doibase 10.1103/PhysRevLett.93.240403} {\bibfield
		{journal} {\bibinfo  {journal} {Phys. Rev. Lett.}\ }\textbf {\bibinfo
			{volume} {93}},\ \bibinfo {pages} {240403} (\bibinfo {year}
		{2004})}\BibitemShut {NoStop}%
	\bibitem [{\citenamefont {Kamchatnov}\ and\ \citenamefont
		{Pitaevskii}(2008)}]{Pitaevskiisnake2008}%
	\BibitemOpen
	\bibfield  {author} {\bibinfo {author} {\bibfnamefont {A.~M.}\ \bibnamefont
			{Kamchatnov}}\ and\ \bibinfo {author} {\bibfnamefont {L.~P.}\ \bibnamefont
			{Pitaevskii}},\ }\href {\doibase 10.1103/PhysRevLett.100.160402} {\bibfield
		{journal} {\bibinfo  {journal} {Phys. Rev. Lett.}\ }\textbf {\bibinfo
			{volume} {100}},\ \bibinfo {pages} {160402} (\bibinfo {year}
		{2008})}\BibitemShut {NoStop}%
	\bibitem [{\citenamefont {Huang}\ \emph {et~al.}(2003)\citenamefont {Huang},
		\citenamefont {Makarov},\ and\ \citenamefont {Velarde}}]{huang2003}%
	\BibitemOpen
	\bibfield  {author} {\bibinfo {author} {\bibfnamefont {G.}~\bibnamefont
			{Huang}}, \bibinfo {author} {\bibfnamefont {V.~A.}\ \bibnamefont {Makarov}},
		\ and\ \bibinfo {author} {\bibfnamefont {M.~G.}\ \bibnamefont {Velarde}},\
	}\href {\doibase 10.1103/PhysRevA.67.023604} {\bibfield  {journal} {\bibinfo
			{journal} {Phys. Rev. A}\ }\textbf {\bibinfo {volume} {67}},\ \bibinfo
		{pages} {023604} (\bibinfo {year} {2003})}\BibitemShut {NoStop}%
	\bibitem [{\citenamefont {Yu}\ and\ \citenamefont {Blakie}(2021)}]{MDWYuBlair}%
	\BibitemOpen
	\bibfield  {author} {\bibinfo {author} {\bibfnamefont {X.}~\bibnamefont
			{Yu}}\ and\ \bibinfo {author} {\bibfnamefont {P.~B.}\ \bibnamefont
			{Blakie}},\ }\href {\doibase 10.1103/PhysRevResearch.3.023043} {\bibfield
		{journal} {\bibinfo  {journal} {Phys. Rev. Research}\ }\textbf {\bibinfo
			{volume} {3}},\ \bibinfo {pages} {023043} (\bibinfo {year}
		{2021})}\BibitemShut {NoStop}%
	\bibitem [{\citenamefont {Yu}\ and\ \citenamefont
		{Blakie}(2022{\natexlab{a}})}]{FDSexact2022}%
	\BibitemOpen
	\bibfield  {author} {\bibinfo {author} {\bibfnamefont {X.}~\bibnamefont
			{Yu}}\ and\ \bibinfo {author} {\bibfnamefont {P.~B.}\ \bibnamefont
			{Blakie}},\ }\href {\doibase 10.1103/PhysRevLett.128.125301} {\bibfield
		{journal} {\bibinfo  {journal} {Phys. Rev. Lett.}\ }\textbf {\bibinfo
			{volume} {128}},\ \bibinfo {pages} {125301} (\bibinfo {year}
		{2022}{\natexlab{a}})}\BibitemShut {NoStop}%
	\bibitem [{\citenamefont {Yu}\ and\ \citenamefont
		{Blakie}(2022{\natexlab{b}})}]{FDScore2022}%
	\BibitemOpen
	\bibfield  {author} {\bibinfo {author} {\bibfnamefont {X.}~\bibnamefont
			{Yu}}\ and\ \bibinfo {author} {\bibfnamefont {P.~B.}\ \bibnamefont
			{Blakie}},\ }\href {\doibase 10.1103/PhysRevResearch.4.033056} {\bibfield
		{journal} {\bibinfo  {journal} {Phys. Rev. Res.}\ }\textbf {\bibinfo {volume}
			{4}},\ \bibinfo {pages} {033056} (\bibinfo {year}
		{2022}{\natexlab{b}})}\BibitemShut {NoStop}%
	\bibitem [{\citenamefont {Stamper-Kurn}\ and\ \citenamefont
		{Ueda}(2013)}]{StamperRMP}%
	\BibitemOpen
	\bibfield  {author} {\bibinfo {author} {\bibfnamefont {D.~M.}\ \bibnamefont
			{Stamper-Kurn}}\ and\ \bibinfo {author} {\bibfnamefont {M.}~\bibnamefont
			{Ueda}},\ }\href {\doibase 10.1103/RevModPhys.85.1191} {\bibfield  {journal}
		{\bibinfo  {journal} {Rev. Mod. Phys.}\ }\textbf {\bibinfo {volume} {85}},\
		\bibinfo {pages} {1191} (\bibinfo {year} {2013})}\BibitemShut {NoStop}%
	\bibitem [{\citenamefont {Kawaguchi}\ and\ \citenamefont
		{Ueda}(2012)}]{KAWAGUCHI12}%
	\BibitemOpen
	\bibfield  {author} {\bibinfo {author} {\bibfnamefont {Y.}~\bibnamefont
			{Kawaguchi}}\ and\ \bibinfo {author} {\bibfnamefont {M.}~\bibnamefont
			{Ueda}},\ }\href {\doibase https://doi.org/10.1016/j.physrep.2012.07.005}
	{\bibfield  {journal} {\bibinfo  {journal} {Phys. Rep.}\ }\textbf {\bibinfo
			{volume} {520}},\ \bibinfo {pages} {253} (\bibinfo {year} {2012})},\ \bibinfo
	{note} {spinor Bose--Einstein condensates}\BibitemShut {NoStop}%
	\bibitem [{\citenamefont {Ho}(1998)}]{Ho98}%
	\BibitemOpen
	\bibfield  {author} {\bibinfo {author} {\bibfnamefont {T.-L.}\ \bibnamefont
			{Ho}},\ }\href {\doibase 10.1103/PhysRevLett.81.742} {\bibfield  {journal}
		{\bibinfo  {journal} {Phys. Rev. Lett.}\ }\textbf {\bibinfo {volume} {81}},\
		\bibinfo {pages} {742} (\bibinfo {year} {1998})}\BibitemShut {NoStop}%
	\bibitem [{\citenamefont {Ohmi}\ and\ \citenamefont {Machida}(1998)}]{OM98}%
	\BibitemOpen
	\bibfield  {author} {\bibinfo {author} {\bibfnamefont {T.}~\bibnamefont
			{Ohmi}}\ and\ \bibinfo {author} {\bibfnamefont {K.}~\bibnamefont {Machida}},\
	}\href {\doibase 10.1143/JPSJ.67.1822} {\bibfield  {journal} {\bibinfo
			{journal} {J. Phys. Soc. Jpn}\ }\textbf {\bibinfo {volume} {67}},\ \bibinfo
		{pages} {1822} (\bibinfo {year} {1998})}\BibitemShut {NoStop}%
	\bibitem [{\citenamefont {Sadler}\ \emph {et~al.}(2006)\citenamefont {Sadler},
		\citenamefont {Higbie}, \citenamefont {Leslie}, \citenamefont
		{Vengalattore},\ and\ \citenamefont {Stamper-Kurn}}]{Stampernatrue2006}%
	\BibitemOpen
	\bibfield  {author} {\bibinfo {author} {\bibfnamefont {L.~E.}\ \bibnamefont
			{Sadler}}, \bibinfo {author} {\bibfnamefont {J.~M.}\ \bibnamefont {Higbie}},
		\bibinfo {author} {\bibfnamefont {S.~R.}\ \bibnamefont {Leslie}}, \bibinfo
		{author} {\bibfnamefont {M.}~\bibnamefont {Vengalattore}}, \ and\ \bibinfo
		{author} {\bibfnamefont {D.~M.}\ \bibnamefont {Stamper-Kurn}},\ }\href
	{https://doi.org/10.1038/nature05094} {\bibfield  {journal} {\bibinfo
			{journal} {Nature}\ }\textbf {\bibinfo {volume} {443}},\ \bibinfo {pages}
		{312} (\bibinfo {year} {2006})}\BibitemShut {NoStop}%
	\bibitem [{foo({\natexlab{a}})}]{footnoteunstable}%
	\BibitemOpen
	\href@noop {} {\emph {\bibinfo {title} {\rm{Note that the unstable modes must
					be localized in space and are insensitive to boundary conditions as long as
					the system size is much larger than $ \ell^{\rm I, \rm II}$ }}}}\BibitemShut
	{NoStop}%
	\bibitem [{spi()}]{spintwist}%
	\BibitemOpen
	\href@noop {} {\emph {\bibinfo {title} {\rm{We emphasize that since the
					unstable modes of spin-twist instabilities keep $F_x$ unchanged, the magnetic
					domain wall structure of both type-I and type-II FDSs do not break down under
					the corresponding perturbations. This is confirmed by numerical simulations
					of real time dynamics of FDS with initially imprinting the unstable modes
					associated with the spin twist instabilities (see Figs.~S3 and~S4~\cite{SM})
	}}}}\BibitemShut {NoStop}%
	\bibitem [{foo({\natexlab{b}})}]{footnotesmallgs}%
	\BibitemOpen
	\href@noop {} {\emph {\bibinfo {title} {\rm{For small spin coupling strength
					$|g_s/g_n|\lesssim0.1$, the spin-twist instability persists in the whole
					easy-plane phase ($0<\tilde{q}<1$)}}}}\BibitemShut {NoStop}%
	\bibitem [{SM()}]{SM}%
	\BibitemOpen
	\href@noop {} {\emph {\bibinfo {title} {\rm{{See Supplemental Material for
						details.}}}}}\BibitemShut {Stop}%
	\bibitem [{foo({\natexlab{c}})}]{footzeroq}%
	\BibitemOpen
	\href@noop {} {\emph {\bibinfo {title} {\rm{When $q \rightarrow 0$, the snake
					instability vanishes [see Fig.~\ref{f:instabilitytypeII}(a) and
					Ref.~\cite{MDWYuBlair}]. In this limit, the system exhibits $\textrm{SO}(3)$
					rotational symmetry and the two types of FDSs are connected by a {\em
						symmetric} unitary spin-1 rotation ($\psi_s^{\rm II}=U \psi_s^{\rm I}$ and
					$U=U^{T}$)~\cite{MDWYuBlair, FDSexact2022} and consequently the two
					spin-twist unstable modes are also related by a spin rotation, namely, $u
					\rightarrow U \, u$ and $v \rightarrow U^{\dagger} \, v$ }}}}\BibitemShut
	{NoStop}%
	\bibitem [{foo({\natexlab{d}})}]{footnotepositivemass}%
	\BibitemOpen
	\href@noop {} {\emph {\bibinfo {title} {\rm{Positive inertial mass does not
					necessarily ensure stable oscillations. This is the case for long phase
					domain walls~\cite{Son2002}, whose inertial mass is positive for weak
					coherent coupling strengths~\cite{MDQu2017, Sophie, Gallemi2019}, which have
					been shown to fragment under evolution~\cite{Gallemi2019,
						Kasamatsu2019}.}}}}\BibitemShut {Stop}%
	\bibitem [{typ()}]{typetransitions}%
	\BibitemOpen
	\href@noop {} {\emph {\bibinfo {title} {\rm{{It is worthwhile mentioning that
						once a hybridized FDS is formed, transition between two types may happen
						through a longitudinal motion of the interface between the two types along
						the soliton line and the condition of reaching the maximum speed is not
						necessarily fulfilled for the transition to occur. As seen in
						Fig.~\ref{f:typeIIphase}(e4), at that time the whole FDS converts to type-I
						and the elements near the locations of the vortices have never moved at the
						speed limit }}}}}\BibitemShut {NoStop}%
	\bibitem [{\citenamefont {Underwood}\ \emph {et~al.}(2023)\citenamefont
		{Underwood}, \citenamefont {Groszek}, \citenamefont {Yu}, \citenamefont
		{Blakie},\ and\ \citenamefont {Williamson}}]{Andrew2023}%
	\BibitemOpen
	\bibfield  {author} {\bibinfo {author} {\bibfnamefont {A.~P.~C.}\
			\bibnamefont {Underwood}}, \bibinfo {author} {\bibfnamefont {A.~J.}\
			\bibnamefont {Groszek}}, \bibinfo {author} {\bibfnamefont {X.}~\bibnamefont
			{Yu}}, \bibinfo {author} {\bibfnamefont {P.~B.}\ \bibnamefont {Blakie}}, \
		and\ \bibinfo {author} {\bibfnamefont {L.~A.}\ \bibnamefont {Williamson}},\
	}\href {\doibase 10.1103/PhysRevResearch.5.L012045} {\bibfield  {journal}
		{\bibinfo  {journal} {Phys. Rev. Res.}\ }\textbf {\bibinfo {volume} {5}},\
		\bibinfo {pages} {L012045} (\bibinfo {year} {2023})}\BibitemShut {NoStop}%
	\bibitem [{foo({\natexlab{e}})}]{footnotefrequency}%
	\BibitemOpen
	\href@noop {} {\emph {\bibinfo {title} {\rm{{The dependence on the inertial
						mass of the oscillation frequency remains an interesting question to resolve,
						especially in the $q\rightarrow0$ limit where the conservation law of
						magnetization may play an important role~\cite{MDWYuBlair}.}}}}}\BibitemShut
	{Stop}%
	\bibitem [{foo({\natexlab{f}})}]{footnotegeneral}%
	\BibitemOpen
	\href@noop {} {\emph {\bibinfo {title} {\rm{{The relevant characteristic
						features of FDSs revealed by the presented results on the strong spin
						interaction regime hold for $0<|g_s/g_n|<1$}}}}}\BibitemShut {NoStop}%
	\bibitem [{\citenamefont {James}\ and\ \citenamefont
		{Lamacraft}(2011)}]{Lamacraft2011}%
	\BibitemOpen
	\bibfield  {author} {\bibinfo {author} {\bibfnamefont {A.~J.~A.}\
			\bibnamefont {James}}\ and\ \bibinfo {author} {\bibfnamefont
			{A.}~\bibnamefont {Lamacraft}},\ }\href {\doibase
		10.1103/PhysRevLett.106.140402} {\bibfield  {journal} {\bibinfo  {journal}
			{Phys. Rev. Lett.}\ }\textbf {\bibinfo {volume} {106}},\ \bibinfo {pages}
		{140402} (\bibinfo {year} {2011})}\BibitemShut {NoStop}%
	\bibitem [{\citenamefont {Son}\ and\ \citenamefont
		{Stephanov}(2002)}]{Son2002}%
	\BibitemOpen
	\bibfield  {author} {\bibinfo {author} {\bibfnamefont {D.~T.}\ \bibnamefont
			{Son}}\ and\ \bibinfo {author} {\bibfnamefont {M.~A.}\ \bibnamefont
			{Stephanov}},\ }\href {\doibase 10.1103/PhysRevA.65.063621} {\bibfield
		{journal} {\bibinfo  {journal} {Phys. Rev. A}\ }\textbf {\bibinfo {volume}
			{65}},\ \bibinfo {pages} {063621} (\bibinfo {year} {2002})}\BibitemShut
	{NoStop}%
	\bibitem [{\citenamefont {Tylutki}\ \emph {et~al.}(2016)\citenamefont
		{Tylutki}, \citenamefont {Pitaevskii}, \citenamefont {Recati},\ and\
		\citenamefont {Stringari}}]{Tylutki2016}%
	\BibitemOpen
	\bibfield  {author} {\bibinfo {author} {\bibfnamefont {M.}~\bibnamefont
			{Tylutki}}, \bibinfo {author} {\bibfnamefont {L.~P.}\ \bibnamefont
			{Pitaevskii}}, \bibinfo {author} {\bibfnamefont {A.}~\bibnamefont {Recati}},
		\ and\ \bibinfo {author} {\bibfnamefont {S.}~\bibnamefont {Stringari}},\
	}\href {\doibase 10.1103/PhysRevA.93.043623} {\bibfield  {journal} {\bibinfo
			{journal} {Phys. Rev. A}\ }\textbf {\bibinfo {volume} {93}},\ \bibinfo
		{pages} {043623} (\bibinfo {year} {2016})}\BibitemShut {NoStop}%
	\bibitem [{\citenamefont {Chomaz}\ \emph {et~al.}(2015)\citenamefont {Chomaz},
		\citenamefont {Corman}, \citenamefont {Bienaim{\'e}}, \citenamefont
		{Desbuquois}, \citenamefont {Weitenberg}, \citenamefont {Nascimb{\`e}ne},
		\citenamefont {Beugnon},\ and\ \citenamefont {Dalibard}}]{Dalibard2015}%
	\BibitemOpen
	\bibfield  {author} {\bibinfo {author} {\bibfnamefont {L.}~\bibnamefont
			{Chomaz}}, \bibinfo {author} {\bibfnamefont {L.}~\bibnamefont {Corman}},
		\bibinfo {author} {\bibfnamefont {T.}~\bibnamefont {Bienaim{\'e}}}, \bibinfo
		{author} {\bibfnamefont {R.}~\bibnamefont {Desbuquois}}, \bibinfo {author}
		{\bibfnamefont {C.}~\bibnamefont {Weitenberg}}, \bibinfo {author}
		{\bibfnamefont {S.}~\bibnamefont {Nascimb{\`e}ne}}, \bibinfo {author}
		{\bibfnamefont {J.}~\bibnamefont {Beugnon}}, \ and\ \bibinfo {author}
		{\bibfnamefont {J.}~\bibnamefont {Dalibard}},\ }\href
	{http://www.nature.com/articles/ncomms7162} {\bibfield  {journal} {\bibinfo
			{journal} {Nat. Commun.}\ }\textbf {\bibinfo {volume} {6}} (\bibinfo {year}
		{2015})}\BibitemShut {NoStop}%
	\bibitem [{\citenamefont {Gauthier}\ \emph {et~al.}(2016)\citenamefont
		{Gauthier}, \citenamefont {Lenton}, \citenamefont {Parry}, \citenamefont
		{Baker}, \citenamefont {Davis}, \citenamefont {Rubinsztein-Dunlop},\ and\
		\citenamefont {Neely}}]{Gauthier16}%
	\BibitemOpen
	\bibfield  {author} {\bibinfo {author} {\bibfnamefont {G.}~\bibnamefont
			{Gauthier}}, \bibinfo {author} {\bibfnamefont {I.}~\bibnamefont {Lenton}},
		\bibinfo {author} {\bibfnamefont {N.~M.}\ \bibnamefont {Parry}}, \bibinfo
		{author} {\bibfnamefont {M.}~\bibnamefont {Baker}}, \bibinfo {author}
		{\bibfnamefont {M.~J.}\ \bibnamefont {Davis}}, \bibinfo {author}
		{\bibfnamefont {H.}~\bibnamefont {Rubinsztein-Dunlop}}, \ and\ \bibinfo
		{author} {\bibfnamefont {T.~W.}\ \bibnamefont {Neely}},\ }\href {\doibase
		10.1364/OPTICA.3.001136} {\bibfield  {journal} {\bibinfo  {journal}
			{\color{blue}Optica}\ }\textbf {\bibinfo {volume} {3}},\ \bibinfo {pages}
		{1136} (\bibinfo {year} {2016})}\BibitemShut {NoStop}%
	\bibitem [{\citenamefont {Semeghini}\ \emph {et~al.}(2018)\citenamefont
		{Semeghini}, \citenamefont {Ferioli}, \citenamefont {Masi}, \citenamefont
		{Mazzinghi}, \citenamefont {Wolswijk}, \citenamefont {Minardi}, \citenamefont
		{Modugno}, \citenamefont {Modugno}, \citenamefont {Inguscio},\ and\
		\citenamefont {Fattori}}]{Semeghini2018}%
	\BibitemOpen
	\bibfield  {author} {\bibinfo {author} {\bibfnamefont {G.}~\bibnamefont
			{Semeghini}}, \bibinfo {author} {\bibfnamefont {G.}~\bibnamefont {Ferioli}},
		\bibinfo {author} {\bibfnamefont {L.}~\bibnamefont {Masi}}, \bibinfo {author}
		{\bibfnamefont {C.}~\bibnamefont {Mazzinghi}}, \bibinfo {author}
		{\bibfnamefont {L.}~\bibnamefont {Wolswijk}}, \bibinfo {author}
		{\bibfnamefont {F.}~\bibnamefont {Minardi}}, \bibinfo {author} {\bibfnamefont
			{M.}~\bibnamefont {Modugno}}, \bibinfo {author} {\bibfnamefont
			{G.}~\bibnamefont {Modugno}}, \bibinfo {author} {\bibfnamefont
			{M.}~\bibnamefont {Inguscio}}, \ and\ \bibinfo {author} {\bibfnamefont
			{M.}~\bibnamefont {Fattori}},\ }\href {\doibase
		10.1103/PhysRevLett.120.235301} {\bibfield  {journal} {\bibinfo  {journal}
			{Phys. Rev. Lett.}\ }\textbf {\bibinfo {volume} {120}},\ \bibinfo {pages}
		{235301} (\bibinfo {year} {2018})}\BibitemShut {NoStop}%
	\bibitem [{\citenamefont {Higbie}\ \emph {et~al.}(2005)\citenamefont {Higbie},
		\citenamefont {Sadler}, \citenamefont {Inouye}, \citenamefont {Chikkatur},
		\citenamefont {Leslie}, \citenamefont {Moore}, \citenamefont {Savalli},\ and\
		\citenamefont {Stamper-Kurn}}]{Higbie2005}%
	\BibitemOpen
	\bibfield  {author} {\bibinfo {author} {\bibfnamefont {J.~M.}\ \bibnamefont
			{Higbie}}, \bibinfo {author} {\bibfnamefont {L.~E.}\ \bibnamefont {Sadler}},
		\bibinfo {author} {\bibfnamefont {S.}~\bibnamefont {Inouye}}, \bibinfo
		{author} {\bibfnamefont {A.~P.}\ \bibnamefont {Chikkatur}}, \bibinfo {author}
		{\bibfnamefont {S.~R.}\ \bibnamefont {Leslie}}, \bibinfo {author}
		{\bibfnamefont {K.~L.}\ \bibnamefont {Moore}}, \bibinfo {author}
		{\bibfnamefont {V.}~\bibnamefont {Savalli}}, \ and\ \bibinfo {author}
		{\bibfnamefont {D.~M.}\ \bibnamefont {Stamper-Kurn}},\ }\href {\doibase
		10.1103/PhysRevLett.95.050401} {\bibfield  {journal} {\bibinfo  {journal}
			{Phys. Rev. Lett.}\ }\textbf {\bibinfo {volume} {95}},\ \bibinfo {pages}
		{050401} (\bibinfo {year} {2005})}\BibitemShut {NoStop}%
	\bibitem [{\citenamefont {Huh}\ \emph {et~al.}(2020)\citenamefont {Huh},
		\citenamefont {Kim}, \citenamefont {Kwon},\ and\ \citenamefont
		{Choi}}]{Huh2020a}%
	\BibitemOpen
	\bibfield  {author} {\bibinfo {author} {\bibfnamefont {S.}~\bibnamefont
			{Huh}}, \bibinfo {author} {\bibfnamefont {K.}~\bibnamefont {Kim}}, \bibinfo
		{author} {\bibfnamefont {K.}~\bibnamefont {Kwon}}, \ and\ \bibinfo {author}
		{\bibfnamefont {J.-y.}\ \bibnamefont {Choi}},\ }\href {\doibase
		10.1103/PhysRevResearch.2.033471} {\bibfield  {journal} {\bibinfo  {journal}
			{Phys. Rev. Research}\ }\textbf {\bibinfo {volume} {2}},\ \bibinfo {pages}
		{033471} (\bibinfo {year} {2020})}\BibitemShut {NoStop}%
	\bibitem [{\citenamefont {Pr{\"u}fer}\ \emph {et~al.}(2022)\citenamefont
		{Pr{\"u}fer}, \citenamefont {Spitz}, \citenamefont {Lannig}, \citenamefont
		{Strobel}, \citenamefont {Berges},\ and\ \citenamefont
		{Oberthaler}}]{prufer2022condensation}%
	\BibitemOpen
	\bibfield  {author} {\bibinfo {author} {\bibfnamefont {M.}~\bibnamefont
			{Pr{\"u}fer}}, \bibinfo {author} {\bibfnamefont {D.}~\bibnamefont {Spitz}},
		\bibinfo {author} {\bibfnamefont {S.}~\bibnamefont {Lannig}}, \bibinfo
		{author} {\bibfnamefont {H.}~\bibnamefont {Strobel}}, \bibinfo {author}
		{\bibfnamefont {J.}~\bibnamefont {Berges}}, \ and\ \bibinfo {author}
		{\bibfnamefont {M.~K.}\ \bibnamefont {Oberthaler}},\ }\href {\doibase
		https://doi.org/10.1038/s41567-022-01779-6} {\bibfield  {journal} {\bibinfo
			{journal} {Nature Physics}\ }\textbf {\bibinfo {volume} {18}},\ \bibinfo
		{pages} {1459} (\bibinfo {year} {2022})}\BibitemShut {NoStop}%
	\bibitem [{\citenamefont {Ihara}\ and\ \citenamefont
		{Kasamatsu}(2019)}]{Kasamatsu2019}%
	\BibitemOpen
	\bibfield  {author} {\bibinfo {author} {\bibfnamefont {K.}~\bibnamefont
			{Ihara}}\ and\ \bibinfo {author} {\bibfnamefont {K.}~\bibnamefont
			{Kasamatsu}},\ }\href {\doibase 10.1103/PhysRevA.100.013630} {\bibfield
		{journal} {\bibinfo  {journal} {Phys. Rev. A}\ }\textbf {\bibinfo {volume}
			{100}},\ \bibinfo {pages} {013630} (\bibinfo {year} {2019})}\BibitemShut
	{NoStop}%
\end{thebibliography}


%

\pagebreak
\widetext

\begin{center}
	\textbf{\large Supplemental Material for `` Absence of the breakdown of ferrodark soltions exhibiting a snake instability ''}
\end{center}

\setcounter{equation}{0}
\setcounter{figure}{0}
\setcounter{table}{0}
\setcounter{page}{1}
\makeatletter
\renewcommand{\theequation}{S\arabic{equation}}
\renewcommand{\thefigure}{S\arabic{figure}}
\renewcommand{\bibnumfmt}[1]{[S#1]}
\renewcommand{\citenumfont}[1]{S#1}

\section{Unstable modes associated with the snake instability}
Unstable modes associated with the snake instability for type-II FDSs generate spatial deformation of $F_x$ and weak  modulation of  $n_{\pm1}$.   The magnetization $F_x$ is always zero at the core and the sign change of $F_{x}$ remains unchanged across the core.  
\begin{figure}[htp] 
	\centering
	\includegraphics[width=0.46\textwidth]{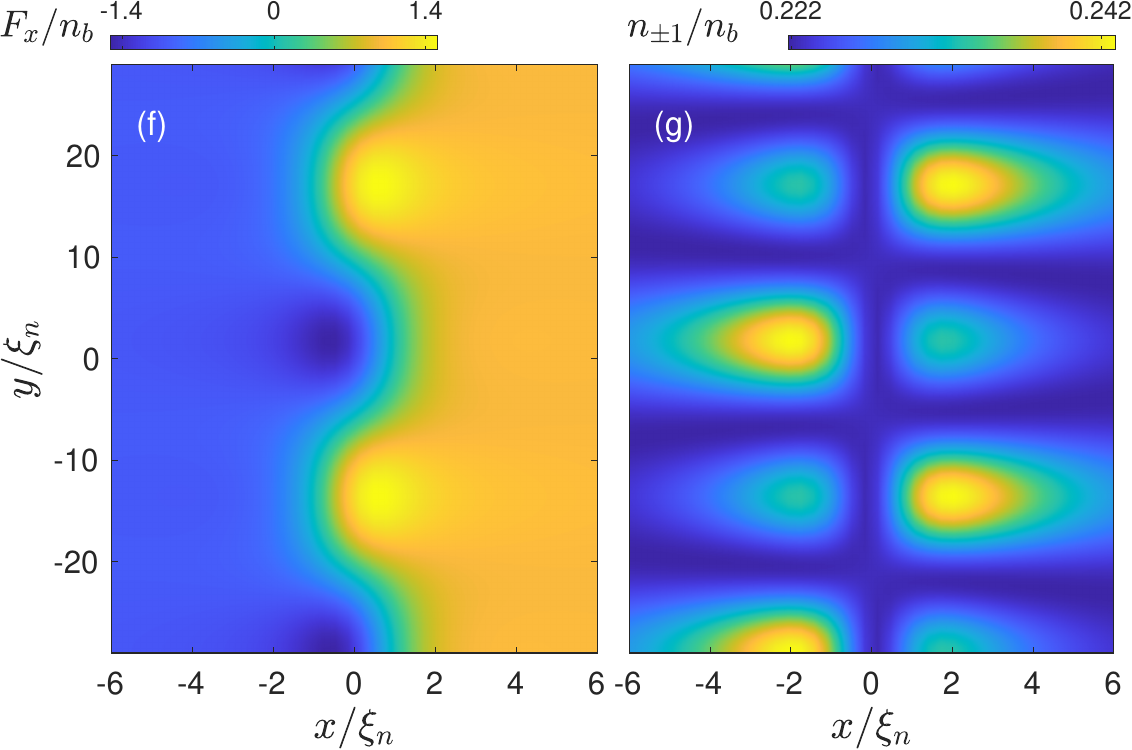}
	\caption{Profiles of $F_x$  and  $n_{\pm 1}$  created by the unstable modes associated with the snake instability.  The unstable modes are chosen  at $k_y \xi_n \simeq 0.205$, $\tilde{q}=0.25$ and we use  $\epsilon=2$.} 
	\label{f:instabilitytypeIImore}
\end{figure}

\section{Evolution of a line type-II FDS at longer times}
\begin{figure}[htp] 
	\centering
	\includegraphics[width=0.46\textwidth]{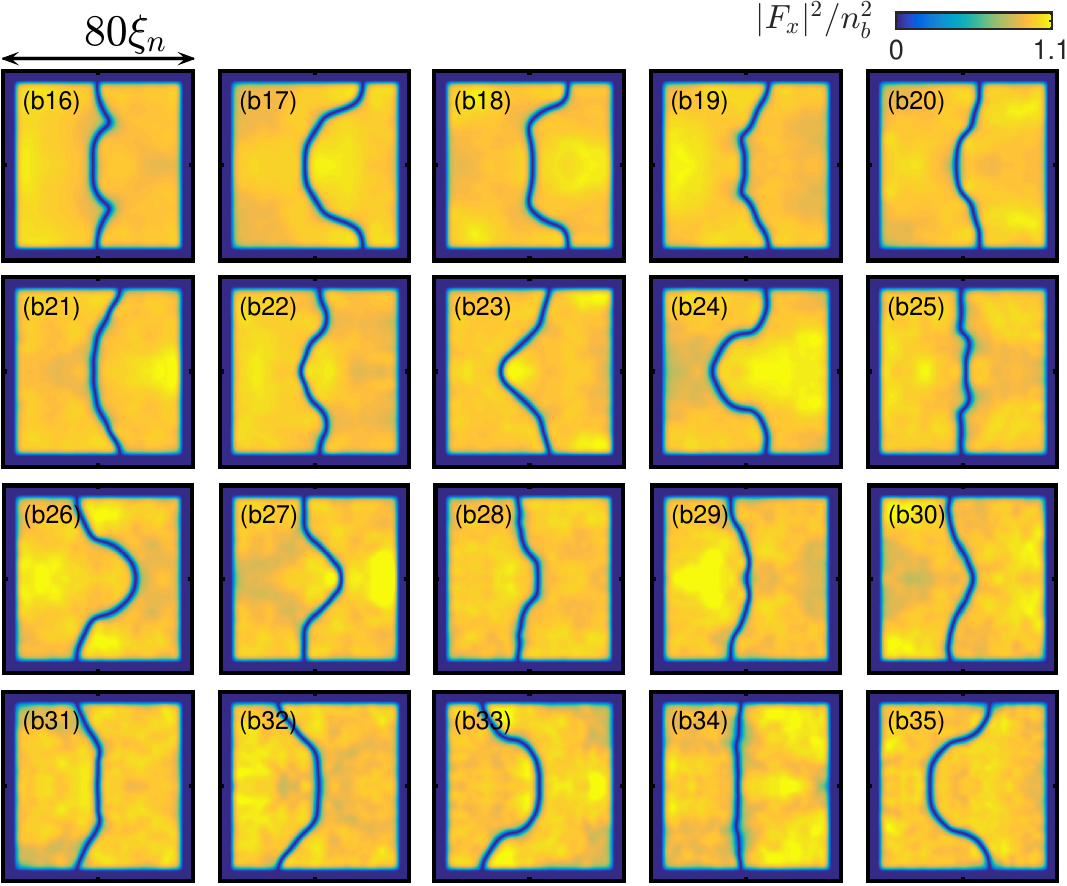}
	\caption{Subsequent evolution of a line type-II FDS after Fig.3 (b15) in the main text  for longer times. The time interval between snapshots is the same as in the main text. } 
	\label{f:typeIIlongtime}
\end{figure}

After the  exponential growth of the amplitude driven by the unstable mode associated with the snake instability, a line  type-II FDS enters the nonlinear regime and forms a vortex-domain wall composite excitation while the topological magnetic domain wall structure persists (Fig.~\ref{f:typeIIlongtime}). 

\section{Dynamics of ferrodark solitons perturbed by the spin-twist unstable modes}

\begin{figure}[htp] 
	\centering
	\includegraphics[width=0.46\textwidth]{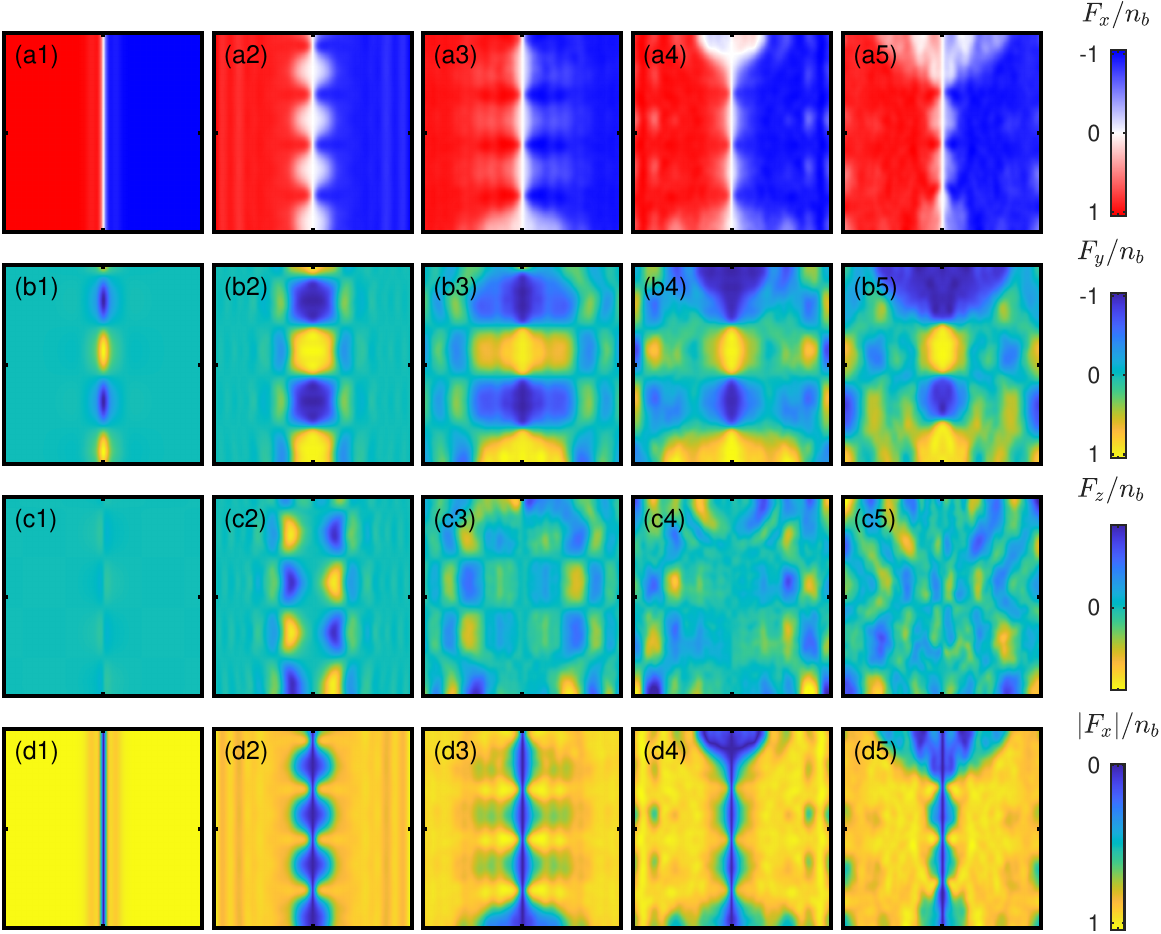}
	\caption{Time evolution of a type-I FDS which is initially perturbed by the unstable mode associated with the type-I spin-twist instability. (a1)-(a5), (b1)-(b5), and (c1)-(c1) show the time evolution of $F_x$, $F_y$ and $F_z$, respectively. (d1)-(d5) The corresponding profile of  $|F_x|$. 
		The system is  in a square domain $x, y \in [-L, L]$ with $L = 40\xi_n$. Here $g_s/g_n=-1/2$,  $\tilde{q}=0.1$, and Neumann boundary condition is applied.  The time interval between snapshots is $40 t_0$. The unstable modes are chosen at $k_y \xi_n \simeq 0.1556$ and we use $\epsilon=0.1$. } 
	\label{f:instabilitytypeItwist}
\end{figure}

\begin{figure}[htp] 
	\centering
	\includegraphics[width=0.46\textwidth]{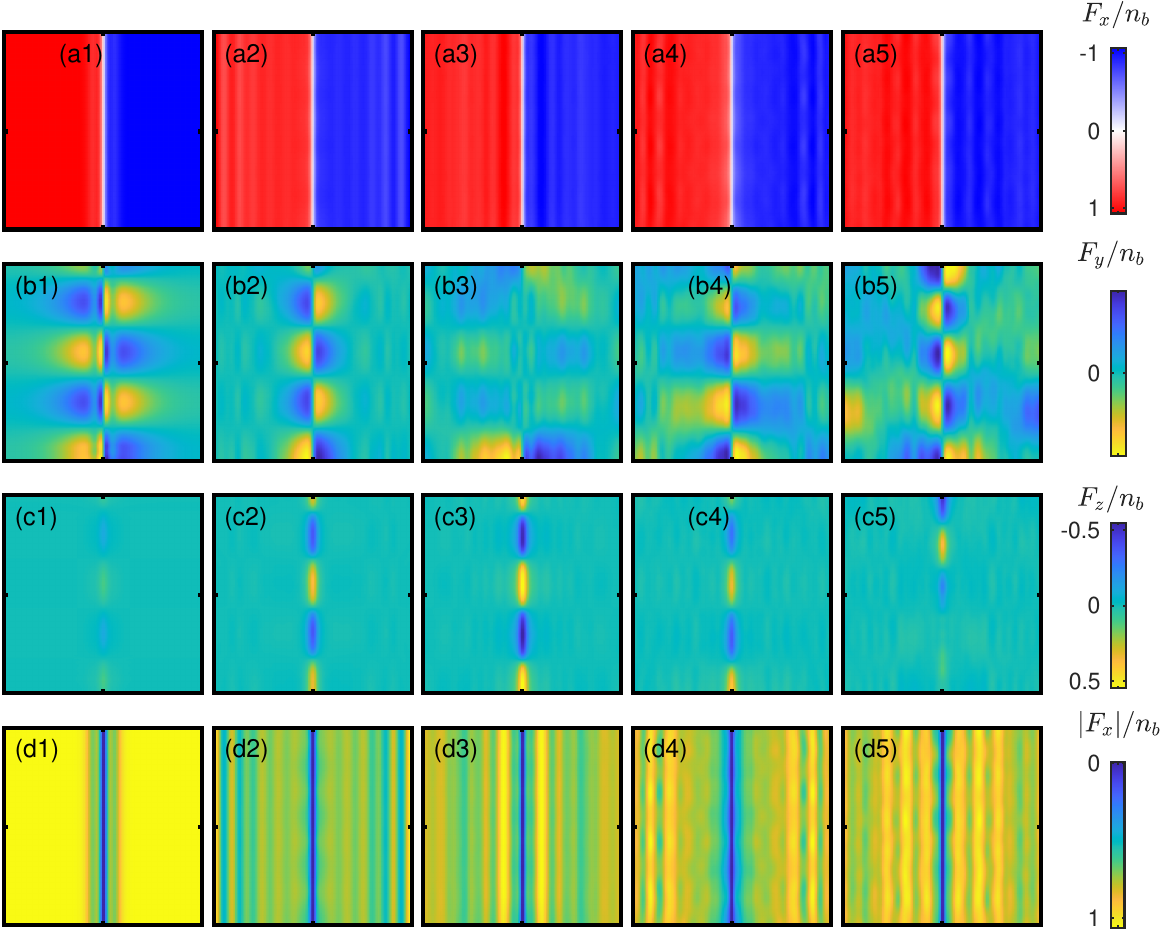}
	\caption{Time evolution of a type-II FDS which is initially perturbed by the unstable mode associated with the type-II spin-twist instability. (a1)-(a5), (b1)-(b5), and (c1)-(c1) show the time evolution of $F_x$, $F_y$ and $F_z$, respectively. (d1)-(d5) The corresponding profile of $|F_x|$. The unstable modes are chosen  at $k_y \xi_n \simeq 0.1556$, $\tilde{q}=0.25$ and we use $\epsilon=0.2$.  The system size, the interaction strength, the time interval between each snapshot, and the boundary condition are the same as in Fig.~\ref{f:instabilitytypeItwist}. } 
	\label{f:instabilitytypeIItwist}
\end{figure}

The initial state is the perturbed  wavefunction $\psi=\psi_s+\delta \psi$, where $\psi_s$ is the wavefunction of a stationary FDS and  $\delta \psi=\epsilon[u(\mathbf{r})e^{-i\omega t}+v^{*}(\mathbf{r})e^{i \omega^{*} t}]$ is the perturbation contributed by the unstable mode associated with the spin-twist instability.  Figures~\ref{f:instabilitytypeItwist} and ~\ref{f:instabilitytypeIItwist}  show the  real time dynamics of a type-I FDS and a type-II FDS, respectively. 

\end{document}